\documentclass{aa}

\usepackage{graphicx}
\usepackage{txfonts}
\usepackage{amsmath}
\usepackage{amssymb}
\usepackage[colorlinks=true,linkcolor=blue,citecolor=blue,urlcolor=blue]{hyperref}
\makeatletter
\renewcommand*\aa@pageof{, page \thepage{} of \pageref*{LastPage}}
\makeatother
\newcommand{\bx}{\mathbf{x}}
\newcommand{\by}{\mathbf{y}}
\newcommand{\bz}{\mathbf{z}}
\newcommand{\bs}{\mathbf{s}}
\newcommand{\bg}{\mathbf{g}}
\newcommand{\ba}{\mathbf{a}}
\newcommand{\bt}{\mathbf{t}}
\newcommand{\bU}{\mathbf{U}}
\newcommand{\bb}{\mathbf{b}}
\newcommand{\mH}{\mathsf{H}}
\newcommand{\mK}{\mathsf{K}}
\newcommand{\dd}{\mathrm{d}}
\newcommand{\nb}{\bar{n}}
\newcommand{\nobs}{\bar n_{\rm obs}}
\newcommand{\nrand}{\bar n_{\rm rand}}
\newcommand{\nfield}{n_{\rm obs}}
\newcommand{\W}{\mathcal{W}}
\newcommand{\Veff}{\mathcal{V}}
\newcommand{\Vtot}{V_{\rm tot}}
\newcommand{\Nb}{N_{\rm b}}
\newcommand{\E}{\mathrm{E}}
\newcommand{\Var}{\mathrm{Var}}
\newcommand{\Cov}{\mathrm{Cov}}
\newcommand{\Shat}{\hat S}
\newcommand{\mA}{\mathsf{A}}
\newcommand{\mR}{\mathsf{R}}
\newcommand{\mM}{\mathsf{M}}
\newcommand{\mJ}{\mathsf{J}}
\newcommand{\bc}{\mathbf{c}}
\newcommand{\bPsi}{\boldsymbol{\Psi}}

\begin{document}

   \title{Clustering leakage into estimated survey selection functions:
          a point-process framework generalising integral constraint}

   \titlerunning{Clustering leakage into estimated selection functions}
   \authorrunning{A. Farina et al.}

   \author{A. Farina\inst{1, 2, 3}
          \and
          A. Veropalumbo\inst{1, 2}
          \and
          B. R. Granett\inst{1}
          \and
          E. Branchini\inst{2, 3, 1}
          }

        \institute{
        INAF - Osservatorio Astronomico di Brera, Via Brera 28, 20122 Milano, Italy; via E. Bianchi 46, 23807 Merate, Italy\\
        \email{antonio.farina@inaf.it}
        \and
        Dipartimento di Fisica, Universit\`a degli Studi di Genova, Via Dodecaneso 33, I-16146 Genova, Italy
        \and
        INFN - Sezione di Genova, Via Dodecaneso 33, I-16146 Genova, Italy
        }

   \date{Received [date]; accepted [date]}

   \abstract
   {The selection function of a galaxy survey is usually estimated
    from the catalogue itself, by comparing the galaxy counts under
    different observing conditions. Because the counts also fluctuate
    with the large-scale structure, the estimate absorbs genuine
    clustering signal.}
   {We aim at a common statistical description of the
    selection-function estimators in use, from binned completeness
    maps to neural networks, and at a framework to quantify the
    clustering they absorb and its effects.}
   {We describe the catalogue as a Poisson point process whose
    intensity follows the cosmological density field, thinned by the
    selection function. A broad class of estimators maximise the
    Poisson likelihood of unclustered galaxies and differ only in how
    the selection function is parametrised. Expanding the likelihood
    to first order in the density fluctuations, we derive, for any
    parametrisation, the clustering absorbed by the estimate and its
    effect on the measured correlation function. We test the results
    for binned maps on 100 simulated halo catalogues with a known
    selection function and on 1000 unclustered control catalogues,
    and compute the effect of the official selection-function weights
    of BOSS DR12 CMASS.}
   {The absorbed clustering is a weighted average of the density
    field. The estimate therefore carries a sample-variance error on
    top of the Poisson one, and the measured correlation function is
    shifted by a generalised integral constraint, linear in the
    correlation function, which can be included in the model. For a
    binned map, each bin absorbs the mean overdensity of the region
    it occupies, and the global integral constraint is the single-bin
    case. On the simulated catalogues both effects, and a further term
    from the Poisson noise of the counts, match the predictions with
    no free parameter. On BOSS CMASS the leakage is negligible at
    baryon-acoustic-oscillation scales but becomes important at the
    largest scales, where the signal is weaker.}
   {}


   \maketitle

\section{Introduction}
\label{sec:intro}

The statistical power of modern galaxy surveys, from spectroscopic
programmes such as BOSS \citep{dawson13}, eBOSS \citep{dawson16}, and
DESI \citep{desi16}, and photometric surveys such as the Dark Energy
Survey \citep{des3yr}, the Kilo-Degree Survey \citep{kids1000}, and
Hyper Suprime-Cam \citep{hikage19}, has pushed statistical errors on
clustering and weak-lensing observables to the percent level. 
The \textit{Euclid} mission \citep{euclid25} and the \textit{Roman}
Space Telescope \citep{spergel15}, which combine both photometry and spectroscopy, will push
them further. At this
precision, systematic errors dominate, and among the most persistent
are those arising from the survey \emph{selection function}, the
spatially varying probability of detecting a galaxy. It depends
on the selection criteria of the survey and on the systematic errors
of the observing strategy and the data analysis, and it has to be
estimated and corrected for before the clustering can be measured.

The standard remedy estimates a selection function $S(\bx)$ and uses
it to build inverse-probability weights \citep{davis82,ross12} or a
weighted random catalogue for pair-count estimators
\citep{ls93,ross12}. In practice $S$ is often obtained by regressing
the observed galaxy density against systematics templates
\citep{ross12,elsner16,elsner17}, by tabulating the data-to-random
ratio in bins of systematics space
\citep{ross11,johnston21,yan25}, or, increasingly,
by learning a non-linear map from observing conditions to expected
counts with machine-learning selection models \citep{rezaie20,rezaie21,chaussidon22,rosadomarin25}. Whatever the
method, $S$ is estimated from the same catalogue whose clustering is
subsequently measured, and this has a cost. Any estimator that
forces the corrected density to match the observed one, averaged over
some region of the survey, removes not only the systematic but also
the true mean overdensity of that region. This loss of signal is
recognised in the literature, in a form specific to each method. In template regression it is the clustering bias that the fitted
templates introduce, for which analytic corrections exist
\citep{elsner16,elsner17}. For flexible machine-learning models it is
usually called overfitting and controlled empirically
\citep{rezaie20,weaverdyck21}. In spectroscopic analyses with weights
fitted on the sky it is an angular integral constraint, estimated by
applying the weighting to mocks \citep{demattia21,chaussidon25,chiarenza26}.

This paper has two aims. The first is a common statistical
description of these estimators. We model the observed catalogue as
a Poisson point process whose intensity follows the cosmological
density field, thinned by the selection function
\citep{daley03,moller04}, and show that a broad class of
estimators of
$S$, from binned completeness maps to
neural-networks, maximise the same likelihood of this
process: the Poisson likelihood one would write if the density
fluctuations were zero. They differ only in how $S$ is
parametrised.

The second aim is to quantify the clustering that the fit absorbs
and removes when the catalogue is corrected. The mechanism is easiest
to see in the simplest estimator, which counts the galaxies found
in bins of an observing condition and compares them with
the area observed under that condition. That area is a set of patches
scattered over the survey. If they happen to be overdense, the excess
of galaxies is attributed to the observing condition, the
completeness is overestimated, and correcting the catalogue by that
estimate removes the mean overdensity of the patches together with
the systematic. We derive this effect on the uncertainty of the
estimate and on the measured correlation function for a general
parametrisation, and test the predictions for this simplest
estimator on simulated halo catalogues with a known selection
function. We then illustrate the size of the effect on real data,
the CMASS sample of BOSS DR12, which requires extending the
treatment to fits that do not maximise the likelihood and to
selection functions that depend on properties of the galaxies.

The paper is structured as follows. Section~\ref{sec:setup}
introduces the point-process model, the selection function, and the
master random catalogue. Section~\ref{sec:lik} derives the likelihood
for a given density field, the likelihood that the estimators
in common use actually maximise, its score, and the $\nb$--$S$
degeneracy. Section~\ref{sec:cases} specialises it to binned
completeness maps and neural-network models.
Section~\ref{sec:variance} derives the bias of the score in a single
realisation and the sandwich covariance of the estimate.
Section~\ref{sec:leakage} derives the leakage into the corrected
field for a general parametrisation, then for binned and
estimated-$\nb$ models, the shot-noise term, and how to compute the
correction in a survey analysis. Section~\ref{sec:valid} tests the
binned-model results on simulated halo catalogues.
Section~\ref{sec:boss} illustrates the effect, as an example, on
the CMASS sample of BOSS DR12.
Section~\ref{sec:conc} summarises the results and discusses the
assumptions and the relation to existing treatments.

\section{Setup and definitions}
\label{sec:setup}

\subsection{The galaxy distribution as a point process}
\label{ssec:pointproc}

A galaxy catalogue is a finite set of points in three-dimensional
space (or angular--redshift space for photometric surveys), naturally
described as a realisation of a \emph{point process} with intensity
$\lambda(\bx)$: $\lambda(\bx)\,\dd\bx$ is the expected number of
galaxies in the infinitesimal volume element $\dd\bx$. The galaxy
distribution is clustered, so it is not a Poisson process, whose
points would be placed independently of one another. The standard
way of describing it \citep{layzer56,peebles80} is to put the
clustering into the intensity and keep the sampling Poisson. One imagines a continuous density field drawn first, and the galaxies
then drawn as a Poisson process from it. We follow this
\emph{Poisson-sampling} model and write the intensity of the true
(unobserved) galaxy distribution, conditional on a given realisation
of the cosmological density field, as
\begin{equation}
   \lambda(\bx) = \nb\,[1+\delta(\bx)]\,,
   \label{eq:lambda_true}
\end{equation}
where $\nb$ is the mean galaxy number density and $\delta(\bx)$ the
cosmological overdensity field, with $\langle\delta\rangle=0$ and
two-point function
$\langle\delta(\bx)\delta(\by)\rangle=\xi(\bx-\by)$. Angle brackets
and $\E_\delta[\cdot]$ both denote this average over realisations of
$\delta$, as distinct from $\E[\cdot\,|\,\delta]$, the Poisson average
at fixed $\delta$. In the point-process literature a Poisson process
whose intensity is itself random is called a \emph{Cox}, or doubly
stochastic, process \citep{cox55,daley03}.

\subsection{Survey systematics and the selection function}

At each position $\bx$ we define a vector $\bs(\bx)$ of
\emph{observational conditions} (seeing, sky background, depth,
airmass, focal-plane position, local stellar density,
spectral-overlap indicators, \ldots) assumed known at every position,
either from the observations themselves or from ancillary data. The
\emph{selection function} is
\begin{equation}
   S(\bs;\,\theta) = P(D=1\,|\,\bs,\,\theta)\,,
   \qquad 0 \le S \le 1\,,
   \label{eq:sf_def}
\end{equation}
where $D \in \{0,1\}$ is a binary detection indicator and $\theta$, a
vector of $P$ parameters, parametrises the dependence of detection
probability on observing conditions. Nothing in Eq.~(\ref{eq:sf_def})
restricts the functional form of $S(\bs;\theta)$. It may be a step
function on bins of $\bs$-space, a linear or log-linear template
regression, or a neural network. We write
$S(\bx;\theta)\equiv S(\bs(\bx);\theta)$ when the dependence on
position is the relevant one.

\subsection{The independence assumption and Poisson thinning}
\label{ssec:thinning}

We work under the \emph{independence assumption}: detection depends
on local observing conditions but not on whether $\bx$ lies in an
overdense or underdense region,
\begin{equation}
   P\!\left(D=1\,\big|\,\bs(\bx),\,\delta(\bx)\right)
   = P\!\left(D=1\,\big|\,\bs(\bx)\right)\,.
   \label{eq:independence}
\end{equation}
This is the standard assumption underlying essentially all current
completeness estimation methods. It is a statement about the
\emph{mechanism}. The systematics maps $\bs(\bx)$ are fixed and do
not respond to $\delta$. It is not a statement that $S(\bx)$ and
$\delta(\bx)$ are uncorrelated over the volume of any single survey. In a finite survey they might be correlated by chance, and
Sect.~\ref{sec:variance} quantifies what such a correlation does to
the estimate. The assumption excludes selection
effects that depend on the local \emph{observed} density: fibre
collisions, blending, and spectral overlap, which lie outside the
present framework
\citep{bianchi17,pinol17,melchior21,euclid26xcii}.

We further assume that each galaxy's detection, given $S$, is an
independent Bernoulli trial. Thinning a Poisson process with a
spatially varying retention probability produces another Poisson
process with intensity rescaled pointwise \citep{daley03}, so thinning
Eq.~(\ref{eq:lambda_true}) gives a Poisson process with intensity
\begin{equation}
   \lambda_{\rm obs}(\bx;\,\theta) =
   \nb\,[1+\delta(\bx)]\,S(\bx;\,\theta)\,,
   \label{eq:lambda_obs}
\end{equation}
exactly, for every realisation of $\delta(\bx)$. This is the
generative model underlying everything that follows. It assumes
that the maps $\bs(\bx)$ record everything the detection probability
depends on. If they do not, for instance because of a calibration
error over some region that no map records, the true selection is
not a function of $\bs$ and cannot be recovered by any $S(\bs;\theta)$.
We do not treat that case.

The selection may also depend on a property of the galaxy,
such as its magnitude or surface brightness, which is correlated with its clustering amplitude. The catalogue is then a
marked point process, and Appendix~\ref{app:marks} shows how the
results below carry over.

\subsection{The master random catalogue}
\label{ssec:master_rand}

All practical calculations are performed with respect to a
\emph{master random catalogue} $\mathcal{R}$: $N_{\rm rand}$ points
drawn uniformly over the survey footprint $\W$, of volume
$\Vtot=\int_\W\dd\bx$, before any completeness weighting, with
uniform density $\nrand = N_{\rm rand}/\Vtot$. Each random point
$a\in\mathcal{R}$ carries the same systematics vector $\bs(a)$ as a
galaxy would at that position.

Every footprint-dependent quantity below (the effective volume, its
gradient, the integral-constraint terms) is an integral over $\W$.
Real footprints are too irregular to integrate analytically, so in
practice these integrals are computed as sums over the randoms. The
tool that justifies this, and that we use repeatedly below, is
Campbell's theorem \citep{campbell09,daley03}: for any point process
with intensity $\lambda(\bx)$ and any function $f(\bx)$,
\begin{equation}
   \E\Bigl[\sum_i f(\bx_i)\Bigr]
   =\int_\W\lambda(\bx)\,f(\bx)\,\dd\bx\,.
   \label{eq:campbell}
\end{equation}
In words, the average of a sum over the points equals the integral
of $f$ weighted by the expected density of points. Only the mean
density enters, so the
theorem holds for any point process, clustered or not. For the master
randoms $\lambda=\nrand$ is constant, and Eq.~(\ref{eq:campbell})
gives
\begin{equation}
   \E\!\left[\sum_{a\in\mathcal{R}}f(\bx_a)\right]
   = \nrand\int_\W f(\bx)\,\dd\bx\,,
   \label{eq:rand_identity}
\end{equation}
which is what justifies replacing each integral by a sum over
$\mathcal R$.

\section{The selection-function likelihood and its score}
\label{sec:lik}

\subsection{The likelihood for a given density field}
\label{ssec:ppp_lik}

Fix a realisation of $\delta$. The field is never observed. We fix it
only as an intermediate step, and every result below depends on it
only through its statistics, $\langle\delta\rangle=0$ and $\xi$. For
a given $\delta$, the detected galaxies are a Poisson process with
intensity
$\lambda_{\rm obs}=\nb(1+\delta)S$ of Eq.~(\ref{eq:lambda_obs}). We
need the probability of observing a given catalogue
$\{\bx_i\}_{i=1}^N$ in $\W$, as a function of $\theta$ and $\nb$. It
follows from the two defining properties of a Poisson process
\citep{daley03}: (i) the number of points in any region $A$ is a
Poisson variable with mean $\int_A\lambda_{\rm obs}\,\dd\bx$; and
(ii) the numbers of points in disjoint regions are independent.

Dividing the footprint into cells small enough to contain at
most one galaxy, multiplying the Poisson probabilities of the occupied
and of the empty cells, and letting the cells shrink
(Appendix~\ref{app:lik}) gives the likelihood
\begin{equation}
   \mathcal{L}(\theta,\nb\,|\,\delta) =
   \exp\!\left[-\int_\W\lambda_{\rm obs}(\bx;\theta)\,\dd\bx\right]
   \prod_{i=1}^{N} \lambda_{\rm obs}(\bx_i;\theta)\,.
   \label{eq:ppp_lik}
\end{equation}
This is the standard Poisson point-process likelihood
\citep{daley03,moller04}. Its two factors have a simple reading. The
product over galaxies is large when the model puts high intensity
where galaxies were actually found. The exponential is the probability
that no galaxy is found anywhere else, and it penalises models that
predict more galaxies in total than were observed.

Substituting $\lambda_{\rm obs}=\nb(1+\delta)S$ and taking the
logarithm,
\begin{align}
   \ln\mathcal{L}(\theta,\nb\,|\,\delta)
   &= N\ln\nb + \sum_i \ln S(\bx_i;\theta) - \nb\,\Veff(\theta)
   \notag\\
   &\quad + \sum_i\ln[1+\delta(\bx_i)]
      - \nb\!\int_\W\!\delta(\bx)\,S(\bx;\theta)\,\dd\bx\,,
   \label{eq:lnL_delta}\\
   \Veff(\theta) &\equiv \int_\W S(\bx;\theta)\,\dd\bx\,.
   \label{eq:Veff}
\end{align}
$\Veff(\theta)$ is the effective survey volume: the volume a complete
($S=1$) survey would need in order to produce the same expected
count. The first line of Eq.~(\ref{eq:lnL_delta}) contains no
$\delta$. The second line collects everything that depends on the
density field.

\subsection{The likelihood actually maximised, and its score}
\label{ssec:composite}

The second line of Eq.~(\ref{eq:lnL_delta}) cannot be evaluated,
because $\delta$ is unknown. Practical analyses therefore drop both terms of
the second line, for different reasons. The first,
$\sum_i\ln[1+\delta(\bx_i)]$, does not depend on $\theta$, so dropping
it is harmless. The second, $\nb\int_\W\delta\,S\,\dd\bx$, does depend
on $\theta$. It is dropped on the assumption that clustering and
systematics are uncorrelated, so that it averages to zero. That holds
on average over realisations of $\delta$, since $\langle\delta\rangle=0$,
but in a finite survey the two can be aligned by chance, and the term
then measures how much the overdensity happens to overlap with the
selection function. After dropping both terms, what remains is
\begin{equation}
   \ln\mathcal{L}(\theta,\nb) =
   N\ln\nb + \sum_i \ln S(\bx_i;\theta) - \nb\,\Veff(\theta)\,,
   \label{eq:composite}
\end{equation}
which is the Poisson likelihood of a process with intensity
$\nb\,S$, i.e. the likelihood one would write if the galaxies were
unclustered. This is the likelihood that a broad class of
methods for estimating the
selection function maximise, as Sect.~\ref{sec:cases} shows. In the
spatial-statistics literature, its score is used as an estimating
function for the intensity of clustered point processes
\citep{schoenberg05,waagepetersen07,waagepetersen09}.

To see what dropping
$\nb\int_\W\delta\,S\,\dd\bx$ costs when the alignment is not zero, we
look at the score $\bU(\theta)=\nabla_\theta\ln\mathcal{L}(\theta,\nb)$,
the gradient of the log-likelihood with respect to the parameters.
The estimate $\hat\theta$ is the point where the score vanishes. If
the score vanishes on average at the true parameters $\theta_0$, the
estimator recovers them. If it does not, the estimate is pushed away
from them. Differentiating Eq.~(\ref{eq:composite}) gives two terms:
the sum over galaxies gives $\sum_i\nabla_\theta\ln S(\bx_i;\theta)$,
and the effective volume gives
$-\nb\nabla_\theta\Veff=-\nb\int_\W\nabla_\theta S\,\dd\bx$. The
logarithmic derivative in the first term recurs throughout the paper,
so we give it a name,
\begin{equation}
   \bg(\bx;\theta)\equiv\nabla_\theta\ln S(\bx;\theta)\,,
   \qquad
   S\,\bg=\nabla_\theta S\,.
   \label{eq:g_def}
\end{equation}
The score is then
\begin{equation}
   \bU(\theta)=\sum_i\bg(\bx_i;\theta)
   -\nb\int_\W\nabla_\theta S(\bx;\theta)\,\dd\bx\,.
   \label{eq:score_def}
\end{equation}
Equation~(\ref{eq:score_def}) was built from the model with
$\delta=0$, but the data are not drawn from that model. For a given $\delta$, the detected galaxies are a Poisson process with intensity
$\lambda_{\rm obs}=\nb(1+\delta)S$ (Eq.~\ref{eq:lambda_obs}). To see
whether the estimator recovers $\theta_0$, we compute the expectation of the score at $\theta_0$ under this correct model. By
Campbell's theorem, with
$\lambda=\lambda_{\rm obs}$ and $f=\bg$, the sum over galaxies has
expectation $\int_\W\nb(1+\delta)\,S\,\bg\,\dd\bx
=\nb\int_\W(1+\delta)\nabla_\theta S\,\dd\bx$. The second term of
Eq.~(\ref{eq:score_def}) cancels its $\delta$-independent part,
leaving
\begin{equation}
   \E[\bU(\theta_0)\,|\,\delta]
   = \nb\!\int_\W\!\delta(\bx)\,\nabla_\theta S(\bx;\theta_0)\,\dd\bx
   \;\equiv\;\bb(\delta)\,.
   \label{eq:score_bias}
\end{equation}
This is exactly the gradient of the term $\nb\int_\W\delta\,S\,\dd\bx$
that was dropped from Eq.~(\ref{eq:lnL_delta}). Evaluated under the
correct model, the score of the reduced likelihood is biased by
precisely the piece of the full likelihood that was discarded. Had
that piece been kept, the score would have zero mean, as the score of
a correctly specified likelihood does.

Averaging $\bb(\delta)$ over realisations of the field,
$\nabla_\theta S$ does not depend on $\delta$ (Sect.~\ref{ssec:thinning})
and $\langle\delta\rangle=0$, so $\E_\delta[\bb]=0$. The score is
unbiased on average, and the estimator is consistent. In any single
survey, however, $\bb(\delta)\neq0$, and the estimate moves to absorb
the part of the overdensity that overlaps with the directions in
which $S$ can change. Sections~\ref{sec:variance} and
\ref{sec:leakage} quantify this shift and show that it is the origin
of an integral constraint-like effect.

\subsection{The \texorpdfstring{$\nb$--$S$}{n-S} degeneracy}
\label{ssec:degeneracy}

Equation~(\ref{eq:composite}) depends on $\nb$ and $\theta$ only
through the product $\phi(\bx)\equiv\nb\,S(\bx;\theta)$. Rescaling
$S\to\alpha S$ and $\nb\to\nb/\alpha$ for any $\alpha>0$ leaves
$\phi$, and hence $\mathcal{L}$, exactly unchanged, whatever the functional
form of $S$. Only the product $\nb S(\bx)$, the observed galaxy
density, is identifiable. Separating $\nb$ from $S$ requires an
external input, such as a measurement of $\nb$ from a complementary
probe or a normalisation convention for $S$, that no manipulation of
$\mathcal{L}$ can supply. 
We therefore fix $\nb$ by convention and estimate $\theta$ by
maximising Eq.~(\ref{eq:composite}) at that value. This choice
is harmless when $S$ has a free overall scale, that is, when its
parameters include one that multiplies $S$ by a constant everywhere.
Since the likelihood depends on $\nb$ and $S$ only through their
product, the data determine $\nb\,\Shat$, and with it the shape of
$\Shat$, but not its level. Maximising Eq.~(\ref{eq:composite}) at
the chosen $\nb$ sets the level so that the expected number of
galaxies matches the observed one, $\nb\,\Veff(\hat\theta)=N$.
One can therefore fit at any convenient value and rescale $\Shat$
afterwards to the convention in use. A common choice for
the fit is the observed number density,
\begin{equation}
   \nobs\equiv N/\Vtot\,,
   \label{eq:nobs}
\end{equation}
which, since the fit enforces $\nb\,\Veff(\hat\theta)=N$, makes
$\Veff(\hat\theta)=\Vtot$. If $\Shat$ is instead used to
downsample the random catalogue, keeping each random with probability
$\Shat$, its values must lie below one. With this convention they do
not, since $\Shat$ has unit mean, so one first divides $\Shat$ by its
maximum over the footprint. Neither normalisation recovers the true
scale of $S$, which is not identifiable, but both give its
\emph{shape}, which is all that a weighted or downsampled random catalogue uses, since the overall number of randoms cancels in the
normalised pair counts.

\section{Selection-function estimators: binned completeness and neural-network models}
\label{sec:cases}

Before turning to the sample-variance analysis we specialise
Eq.~(\ref{eq:composite}) to two estimators in current use. The
first, the binned completeness map, is the simplest case and will
serve as the worked example for every result of
Sects.~\ref{sec:variance}--\ref{sec:leakage}, where it reduces the
general formulae to region averages of $\delta$ and $\xi$, and
is the model tested in Sect.~\ref{sec:valid}. The
second, a neural-network model, is the most flexible, and maximises
Eq.~(\ref{eq:composite}) directly on the catalogues.

\subsection{The step-function model}
\label{ssec:stepfn}

The simplest and most widely used parametrisation assigns each bin of
systematics space its own free completeness value. It is the model
implicitly assumed whenever the galaxy density is corrected by its
mean in bins of one or more observing conditions. Divide systematics space into $\Nb$ bins $\{B_k\}$ and take
$S(\bs)=p_k$ for $\bs\in B_k$. On the sky, bin $k$ occupies the region
\begin{equation}
   A_k \equiv \{\bx\in\W : \bs(\bx)\in B_k\}\,,
   \qquad V_k=\int_{A_k}\dd\bx\,,
   \label{eq:Ak}
\end{equation}
which is in general a scattered collection of patches (all the
places where the seeing was in a given range, say) rather than a
contiguous area. With the random-catalogue volume estimate
$V_k=N_{{\rm rand},k}/\nrand$, where $N_{{\rm rand},k}$ is the number
of randoms in $A_k$, and $N_k$ galaxies in bin $k$,
Eq.~(\ref{eq:composite}) becomes
\begin{equation}
   \ln\mathcal{L}(\{p_k\}) =
   \sum_k N_k\ln p_k - \nb\sum_k p_k V_k + \mathrm{const}\,,
   \label{eq:step_lik}
\end{equation}
whose maximum is at $N_k/p_k=\nb V_k$ for each bin separately, i.e.
\begin{equation}
   \hat{p}_k = \frac{N_k}{\nb\,V_k}
   = \frac{\nrand}{\nb}\,\frac{N_k}{N_{{\rm rand},k}}\,.
   \label{eq:pk_mle}
\end{equation}
With $\nb=\nobs$ (Sect.~\ref{ssec:degeneracy}) this becomes
\begin{equation}
   \hat p_k=\frac{N_{\rm rand}}{N}\,\frac{N_k}{N_{{\rm rand},k}}\,,
   \label{eq:pk_nobs}
\end{equation}
the fraction of galaxies over the fraction of randoms in the bin:
the relative completeness of Sect.~\ref{ssec:degeneracy}, with unit
mean over the survey volume, which is what a survey can estimate.

\subsection{Extension to neural-network selection models}
\label{ssec:nn}

Take $S(\bs;\theta)$ to be a flexible, unbinned function of the
systematics, such as a neural network with weights $\theta$, evaluated
directly on the catalogue entries. The overall scale of such a model is usually not a
free parameter, since the output range is fixed by the architecture,
and fixing $\nb$ also forces the output to reach a particular
normalisation, which a bounded output (e.g. a sigmoid) may not be
able to reach. This is avoided by removing $\nb$ from the problem
altogether. Replacing $\nb$ in
Eq.~(\ref{eq:composite}) by its best-fit value at fixed $\theta$,
$N/\Veff(\theta)$, and evaluating $\Veff$ with the randoms,
$\Veff(\theta)\simeq\nrand^{-1}\sum_{a\in\mathcal R}S(\bx_a;\theta)$,
gives, up to a $\theta$-independent constant, minus the
log-likelihood. We call it the loss, to be minimised,
\begin{equation}
   \mathcal{J}(\theta)=
   -\sum_{i=1}^N \ln S(\bx_i;\theta)
   + N\ln\,\Bigl[\sum_{a\in\mathcal R}S(\bx_a;\theta)\Bigr]\,,
   \label{eq:nn_loss}
\end{equation}
in which only the systematics columns of the two catalogues enter.
It is unchanged under $S\to\alpha S$, so it needs
no convention for $\nb$, and its minima are the maxima of
Eq.~(\ref{eq:composite}) up to the normalisation of $S$. Equation~(\ref{eq:nn_loss})
is minus the log-probability of the observed galaxy positions under
the normalised density $S/\int_\W S\,\dd\bx$, with the normalisation
estimated on the randoms. Without the second term, the loss would be
minimised by making $S$ as large as possible everywhere. With it,
raising $S$ everywhere gains nothing, and only the shape of $S$ is
fitted. Because the second term is the
logarithm of a sum over the whole random catalogue, estimating it
from minibatches is biased (the logarithm of a mean is not the mean
of the logarithms), and the usual care in stochastic optimisation
applies \citep{belghazi18}.

Networks trained on pixelised maps instead of catalogues minimise the Poisson negative log-likelihood
$\sum_p[\lambda_p-n_p\ln\lambda_p]$, with $n_p$ the galaxy count and
$\lambda_p$ the model expectation in pixel $p$
\citep{rezaie21,rosadomarin25}. With $\lambda_p=\nb\,S_p\,\Delta V_p$ this is
Eq.~(\ref{eq:composite}) with the galaxies and the footprint binned
into pixels. Equation~(\ref{eq:nn_loss}) is the same likelihood
without the pixelisation and with $\nb$ profiled out, so $S$ is
evaluated at the observing conditions at the position of each object
rather than at their average over a pixel. Earlier versions used a
mean-squared error \citep{rezaie20}. Such a fit does not maximise
Eq.~(\ref{eq:composite}), but it still responds linearly to the data,
and the results of Sects.~\ref{sec:variance} and \ref{sec:leakage}
carry over with the weights of its estimating equation
(Appendix~\ref{app:ee}). The density pattern it absorbs is then a
differently weighted average of $\delta$.

The results of Sects.~\ref{sec:variance} and \ref{sec:leakage} apply
to a network, with $\bg$ the gradient of $\ln S$ with respect to its
weights, provided the trained
network responds linearly to small changes in the data. If the
training is regularised, the response is set by the curvature of the
regularised loss rather than by $\mH$ alone, as discussed in
Appendix~\ref{app:reg}. We do not
test this case.

\section{Sample variance of the selection-function estimate}
\label{sec:variance}

In a single survey the density field does not average away, and the
score of Eq.~(\ref{eq:score_def}) is biased by the term of
Eq.~(\ref{eq:score_bias}). This section derives what that bias does
to the estimate of the selection function: by how much the fitted
parameters are displaced in a given realisation, and how large the
resulting scatter is over realisations. We write $\theta_0$ for the
true parameters, so that $S(\bx;\theta_0)$ is the true selection
function, keep $\nb$ fixed as in Sect.~\ref{ssec:degeneracy}, and
work to first order in $\delta$ and in the displacement
$\Delta\theta\equiv\hat\theta-\theta_0$.

\label{ssec:score}
The estimate $\hat\theta$ is the point where the score vanishes,
$\bU(\hat\theta)=0$. To see how far it lies from $\theta_0$, we
expand the score around the truth. Minus its slope there, averaged
at $\delta=0$, is the Fisher information matrix $\mH$ of the
Poisson likelihood, which measures how sharply the galaxy counts
constrain $\theta$. Using $S\nabla\nabla^{\!T}\ln S=\nabla\nabla^{\!T}S-\nabla
S\nabla^{\!T}S/S$ and Campbell's theorem, (Eq.~\ref{eq:campbell}),
the second-derivative terms cancel and
\begin{equation}
   \mH = \nb\int_\W\frac{\nabla_\theta S\,\nabla_\theta^{T}S}{S}\,\dd\bx
       = \nb\int_\W S\,\bg\,\bg^{T}\,\dd\bx\,.
   \label{eq:fisher}
\end{equation}
To first order, $0=\bU(\hat\theta)\simeq\bU(\theta_0)-\mH\,\Delta\theta$,
so the displacement of the estimate is the score at the truth
divided by its slope,
\begin{equation}
   \Delta\theta \simeq \mH^{-1}\bU(\theta_0)
   = \mH^{-1}\bb(\delta) + \mH^{-1}\bU_{\rm P}\,.
   \label{eq:dtheta}
\end{equation}
The score at the truth has been split into its mean at fixed
density field, $\bb(\delta)$ of Eq.~(\ref{eq:score_bias}), and the
Poisson fluctuation around it, $\bU_{\rm P}\equiv\bU(\theta_0)-\bb(\delta)$.
The second term is the familiar shot-noise error of the fit, which
shrinks as the number of galaxies grows. The first does not shrink, because it comes from the density field of the particular survey, not from
the sampling, so it remains however many galaxies are observed. It
is a sample-variance error, a fixed shift of the estimate determined
by the overdensity, averaged over the footprint with weights
$\nabla_\theta S$,
\begin{equation}
   \Delta\theta_{\rm SV} = \mH^{-1}\,\nb\int_\W\delta(\bz)\,\nabla_\theta S(\bz;\theta_0)\,\dd\bz\,.
   \label{eq:dtheta_sv}
\end{equation}
In other words, the fit cannot tell a density fluctuation that happens to follow
$\nabla_\theta S$ from a change of the selection function, and
absorbs it.

\label{ssec:sandwich}
The displacement of Eq.~(\ref{eq:dtheta}) differs from survey to
survey, so it adds to the scatter of the estimate. Being the score at
the true parameters multiplied by $\mH^{-1}$, its covariance is the
variance of that score, multiplied by $\mH^{-1}$ on both sides. The
score fluctuates both through the Poisson sampling at fixed density
field and through the variation of the density field between
realisations. By the law of total covariance, its variance is the sum
of the mean conditional Poisson covariance and the covariance of its
conditional mean. The first is the
Poisson scatter of the galaxies around a given density field, whose
variance, averaged over realisations of the field, is exactly $\mH$. The second is
the scatter of $\bb(\delta)$ from one density field to
another,
\begin{equation}
   \mK \equiv \Var_\delta(\bb)
   = \nb^2\iint_\W \nabla_\theta S(\bx)\,\nabla_\theta^{T}S(\by)\,
     \xi(\bx-\by)\,\dd\bx\,\dd\by\,.
   \label{eq:K}
\end{equation}
Hence, with the linearisation of Eq.~(\ref{eq:dtheta}) as the only
approximation,
\begin{equation}
   \Cov(\hat\theta) \simeq \mH^{-1} + \mH^{-1}\mK\,\mH^{-1}\,.
   \label{eq:sandwich}
\end{equation}
This is the sandwich covariance of
intensity-parameter estimates in clustered point processes, derived
by \citet{waagepetersen07} for inhomogeneous Neyman--Scott processes
\citep{neyman58}
and by \citet{waagepetersen09} for general inhomogeneous processes,
so called because the clustering term $\mK$ is sandwiched between
two inverse Fisher information matrices. Here the clustering is that of the
cosmological $\xi$. $\mK$ can be computed from a model $\xi$ with
the random catalogue, since its integrals are of integral-constraint
type (Sect.~\ref{sec:leakage}), or from mocks.

The binned model of Sect.~\ref{ssec:stepfn} makes all of this
explicit. Changing $p_k$ changes $S$ only inside the region $A_k$, so
the gradient of $S$ with respect to $p_k$ is the indicator of $A_k$,
$\nabla_{p_k}S=\mathbf 1_{A_k}$,
and the quantities above become
\begin{equation}
   \mH_{kl}=\frac{\nb V_k}{p_k}\,\delta^{\rm K}_{kl}\,,\qquad
   \bb_k=\nb V_k\,\bar\delta_k\,,\qquad
   \mK_{kl}=\nb^2V_kV_l\,\bar\xi_{kl}\,,
   \label{eq:HbK_binned}
\end{equation}
where $\delta^{\rm K}_{kl}$ is the Kronecker delta ($\mH$ is
diagonal because the regions $A_k$ do not overlap),
$\bar\delta_k$ is the mean of $\delta$ over $A_k$, and
\begin{equation}
   \bar\xi_{kl}\equiv\frac{1}{V_kV_l}\int_{A_k}\!\dd\bx\int_{A_l}\!\dd\by\;\xi(\bx-\by)
   \label{eq:xibar_kl}
\end{equation}
the correlation function averaged over both regions, equal to
$\Cov(\bar\delta_k,\bar\delta_l)$. Equation~(\ref{eq:dtheta_sv}) then
gives, for the expectation at fixed density field,
\begin{equation}
   \E[\hat p_k\,|\,\delta] = p_k\,(1+\bar\delta_k)\,,
   \label{eq:pk_sv}
\end{equation}
that is, at fixed density field, the expected completeness estimate absorbs the mean overdensity of
the region occupied by its bin. The individual estimate scatters
around this value by its shot noise. Analogously, Eq.~(\ref{eq:sandwich}) gives
\begin{equation}
   \frac{\Cov(\hat p_k,\hat p_l)}{p_k\,p_l}
   = \frac{\delta^{\rm K}_{kl}}{\nb\,p_k\,V_k} + \bar\xi_{kl}\,.
   \label{eq:pk_var}
\end{equation}
The first term is the Poisson error of each bin, one over the
expected number of galaxies in it. It appears only on the diagonal,
since different bins are counted from different galaxies. The second
is the covariance of the mean overdensities of the two regions, and
on the diagonal the variance of the mean overdensity of $A_k$, the sample-variance error.

For the binned model these results are exact, not only first order
in $\delta$, as long as $\nb$ is fixed by convention. The estimate of Eq.~(\ref{eq:pk_mle}) is a count divided by a constant, so
Eqs.~(\ref{eq:pk_sv}) and (\ref{eq:pk_var}) are the mean and variance
of a Poisson count in a given density field, with no expansion
involved.

The size of the sample-variance term depends on how the bins are
chosen. Finer bins in systematics space give smaller regions, whose
mean overdensity fluctuates more, so $\bar\xi_{kk}$ usually grows.
The exception is a region made of many distant patches, whose
overdensities partly average out. With fine bins the completeness estimate becomes in general noisier not only through shot noise but,
increasingly, through the large-scale structure it absorbs.

\section{Leakage into the corrected field}
\label{sec:leakage}

\subsection{General parametrisation: the corrected field and its two-point function}
\label{ssec:leak_general}

Section~\ref{sec:variance} gave the displacement $\Delta\theta_{\rm SV}$
of the selection-function parameters in one realisation. We now
propagate it to the quantity the survey actually measures. The
clustering analysis corrects the observed density for the estimated
selection function,
\begin{equation}
   1+\hat\delta(\bx) = \frac{\nfield(\bx)}{\nb\,\Shat(\bx)}\,,
   \qquad \Shat(\bx)\equiv S(\bx;\hat\theta)\,,
   \label{eq:delta_hat_def}
\end{equation}
where $\nfield(\bx)=\sum_i\delta_{\rm D}(\bx-\bx_i)$ is the observed
number density (a field, not to be confused with its mean $\nobs$),
and the correction is applied either explicitly through
inverse-probability weights $1/\Shat$ on the galaxies or implicitly
by weighting the random catalogue by $\Shat$. To first order in
$\Delta\theta$ the fitted selection function differs from the true
one by the fraction $\bg(\bx)^T\Delta\theta$, and since
$\nfield=\nb(1+\delta)S$ up to shot noise, expanding to first
order in $\delta$ and $\Delta\theta$ (the product
$\delta\,\bg^T\Delta\theta$ is second order and is dropped) and
leaving the shot noise aside, to be treated in
Sect.~\ref{ssec:shotnoise}, the corrected field is
$\hat\delta(\bx)\simeq\delta(\bx)-\bg(\bx)^T\Delta\theta$. The sensitivity $\bg$ converts an error in the parameters into an error
in $\hat\delta$. Inserting the sample-variance part of $\Delta\theta$
from Eq.~(\ref{eq:dtheta_sv}),
\begin{equation}
   \hat\delta(\bx) \simeq \delta(\bx)
   - \bg(\bx)^{T}\mH^{-1}\,\nb\!\int_\W\!\nabla_\theta S(\bz)\,\delta(\bz)\,\dd\bz\,.
   \label{eq:delta_hat}
\end{equation}
The corrected field is the true field minus a specific linear
functional of itself: the component that the selection-function
estimate was able to absorb. Its two-point function involves the
cross-covariance between the score displacement $\bb(\delta)$ and the
density at a second point $\by$, which we define as the vector field
\begin{equation}
   \ba(\by) \equiv \nb\int_\W\nabla_\theta S(\bz)\,\xi(\bz-\by)\,\dd\bz
   = \langle\bb(\delta)\,\delta(\by)\rangle\,.
   \label{eq:a_def}
\end{equation}
Then $\langle\hat\delta(\bx)\hat\delta(\by)\rangle=\xi(\bx-\by)+\Delta\xi(\bx,\by)$
with
\begin{equation}
   \begin{aligned}
   \Delta\xi(\bx,\by) ={}&
   -\,\bg(\bx)^{T}\mH^{-1}\ba(\by)
   -\,\bg(\by)^{T}\mH^{-1}\ba(\bx)\\
   &+\,\bg(\bx)^{T}\mH^{-1}\mK\,\mH^{-1}\bg(\by)\,.
   \end{aligned}
   \label{eq:dxi}
\end{equation}
The selection model enters Eq.~(\ref{eq:dxi}) only through $S$ and
$\nabla_\theta S$, that is, through the density patterns the fit can
absorb. The result therefore holds for any parametrisation of $S$, to
first order in $\delta$ and $\Delta\theta$, and is linear in $\xi$.
The first two terms subtract the cross-correlation between the
true field and the component absorbed by the fit, once for each
point. The last adds the autocorrelation of the absorbed component.
Together they are the two-point function of the residual field of
Eq.~(\ref{eq:delta_hat}).
Equation~(\ref{eq:dxi}) is the main result of this paper. A
pair-count estimator in a separation bin $r$ measures the
pair-weighted average
\begin{equation}
   \langle\Delta\xi\rangle_r =
   \frac{\iint_{|\bx-\by|\in r} w(\bx,\by)\,\Delta\xi(\bx,\by)\,\dd\bx\,\dd\by}
        {\iint_{|\bx-\by|\in r} w(\bx,\by)\,\dd\bx\,\dd\by}\,,
   \label{eq:dxi_r}
\end{equation}
where $w(\bx,\by)$ is the weight with which a pair at $(\bx,\by)$
enters the estimator, i.e. its share of the (weighted) pair counts in
the bin. It depends on how the selection function is applied. If the
galaxies are weighted by $1/\Shat$ and the randoms are uniform, all
parts of the footprint contribute equally and $w=1$. If the galaxies
are unweighted and the randoms are weighted or downsampled by $\Shat$,
regions of low completeness contribute fewer pairs and
$w=S(\bx)S(\by)$.

\subsection{Binned completeness: one integral constraint per bin}
\label{ssec:leak_binned}

For the binned model the result can be seen without the general
formula. With $\nb$ fixed, inserting the estimate of Eq.~(\ref{eq:pk_mle}) in
the corrected field of Eq.~(\ref{eq:delta_hat_def}) gives, for
$\bx\in A_k$,
\begin{equation}
   1+\hat\delta(\bx) = \frac{V_k}{N_k}\,\nfield(\bx)\,.
   \label{eq:deltahat_binned}
\end{equation}
In each region the corrected field is the observed density normalised
to its own mean there. Each bin therefore imposes an integral constraint on the corrected
field, which has mean exactly zero over the region the bin occupies,
in every realisation. Since $N_k$ is proportional to
$1+\bar\delta_k$, to first order Eq.~(\ref{eq:deltahat_binned}) is
$\hat\delta=\delta-\bar\delta_k$ in $A_k$. The fit removes the mean
overdensity of each region.

The general formula reproduces this. Inserting the binned $\mH$,
$\bg$, and $\mK$ of Eq.~(\ref{eq:HbK_binned}) and
$\ba_k(\by)=\nb V_k\,\bar\xi_k(\by)$, where
\begin{equation}
   \bar\xi_k(\by)\equiv\frac{1}{V_k}\int_{A_k}\xi(\bz-\by)\,\dd\bz
   \label{eq:xibar_k}
\end{equation}
is the mean correlation between the point $\by$ and the region
$A_k$, Eq.~(\ref{eq:dxi}) becomes, for $\bx\in A_k$ and $\by\in A_l$,
\begin{equation}
   \Delta\xi(\bx,\by) = -\,\bar\xi_k(\by) - \bar\xi_l(\bx) + \bar\xi_{kl}\,.
   \label{eq:dxi_binned}
\end{equation}
This is the correlation of $\delta-\bar\delta_k$ with
$\delta-\bar\delta_l$, as the direct argument above implies. Each point loses its mean correlation with the region of the other, and
the mean correlation between the two regions, $\bar\xi_{kl}$ of
Eq.~(\ref{eq:xibar_kl}), which is $\bar\xi_k$ further averaged over
$A_l$, is added back. The leakage thus depends on the geometry of
the regions, including their disconnected patches, and not on the
completeness values assigned to them.

The single-bin case recovers the global integral constraint. With
one bin covering the whole survey, $A_1=\W$, the fit only sets the
overall normalisation, and Eq.~(\ref{eq:dxi_binned}) is the
correlation of $\delta$ with its survey mean removed. Writing $\bar\xi_\W$ and $\bar\xi_{\W\W}$ for
the averages of Eqs.~(\ref{eq:xibar_k}) and (\ref{eq:xibar_kl}) over
the whole survey, its average over the pairs in a separation bin is
\begin{equation}
   \langle\Delta\xi\rangle_r = -2\,\langle\bar\xi_\W\rangle_r + \bar\xi_{\W\W}\,,
   \label{eq:ic_single}
\end{equation}
where $\langle\bar\xi_\W\rangle_r$ averages $\bar\xi_\W(\bx)$ over
the pairs at separation $r$. At separations small compared with the
survey the pairs sample it uniformly, so
$\langle\bar\xi_\W\rangle_r\to\bar\xi_{\W\W}$ and the correction is
the constant $-\bar\xi_{\W\W}$, the global integral constraint
\citep{groth77,peebles80,ls93}. At larger $r$ it depends on $r$.

The regions $A_k$ on which the constraints act follow the pattern of the systematic on the
sky, and are in general scattered over the whole survey. Finer bins
give smaller regions and, unless a region is split into many distant
patches, a larger $\bar\xi_{kk}$, so the leakage grows. Equation~(\ref{eq:dxi_binned}) is the
integral-constraint formalism of \citet{demattia19} applied to the
bins of a systematic. Their radial and angular constraints
correspond to redshift shells and sky pixels.

\label{ssec:profiled_nb}
If $\nb$ is also estimated from the data, nothing changes for
parametrisations with a free overall scale, such as the binned one, since $\nb$ is degenerate with the scale (Sect.~\ref{ssec:degeneracy}) and the global constraint is already contained in Eq.~(\ref{eq:dxi}).
For a parametrisation without a free scale, such as a network with a
bounded output, fitting $\nb$ lets the fit rescale $S$ by the same
factor everywhere, which the network cannot do on its own.
Treating $\nb$ as one more parameter, with its own entry in $\bg$,
Eq.~(\ref{eq:dxi}) acquires the $S$-weighted global integral
constraint. Profiling $\nb$ out, as in Eq.~(\ref{eq:nn_loss}), gives
the same result.

\subsection{The shot-noise cross term}
\label{ssec:shotnoise}

Estimating the completeness from the catalogue also absorbs part
of its Poisson noise, which produces a term at non-zero separation,
distinct from the clustering leakage. A random excess of
galaxies in some region raises the fitted completeness there, which
then weights those same galaxies down. Because the completeness is estimated from the galaxies whose pairs are then counted, their shot
noise is partly removed by the fit. This term comes from the
Poisson part of $\Delta\theta$, $\mH^{-1}\bU_{\rm P}$ in
Eq.~(\ref{eq:dtheta}), which Eq.~(\ref{eq:delta_hat}) left out. Writing the shot noise of the observed
density as $\epsilon(\bx)$, a white noise with variance $\nb S$, the
Poisson part of the corrected field has two pieces,
\begin{equation}
   \hat\delta_{\rm P}(\bx)=\frac{\epsilon(\bx)}{\nb S(\bx)}
   -\bg(\bx)^T\mH^{-1}\!\int_\W\bg\,\epsilon\,\dd\bz\,,
   \label{eq:delta_hat_P}
\end{equation}
the shot noise of the galaxies themselves and the shot noise
transmitted through the fit. Its contribution to the two-point
function follows by correlating Eq.~(\ref{eq:delta_hat_P}) at two
distinct points $\bx\neq\by$. The first piece is uncorrelated
between them and drops out, but it is correlated with the second,
because the fit is made with the same galaxies. This correlation,
and not the uncertainty of $\Shat$ by itself, gives
\begin{equation}
   \Delta\xi_{\rm P}(\bx,\by) = -\,\bg(\bx)^{T}\mH^{-1}\bg(\by)\,,
   \qquad \bx\neq\by\,,
   \label{eq:dxi_poisson}
\end{equation}
to be added to Eq.~(\ref{eq:dxi}).

For the binned model the shot-noise term of Eq.~(\ref{eq:dxi_poisson})
is $-1/\E[N_k]$ for two galaxies in the same bin $k$, and zero for
galaxies in different bins.
Averaged over the pairs at separation $r$,
\begin{equation}
   \langle\Delta\xi_{\rm P}\rangle_r = -\sum_k\frac{f_{kk}(r)}{\E[N_k]}\,,
   \label{eq:dxi_poisson_binned}
\end{equation}
where $f_{kk}(r)$ is the geometric fraction of pairs at separation $r$
with both members in bin $k$, counted on the master randoms with
the pair weights of the estimator. The term is largest at small separations, where most
pairs fall within a single region, and fades where the bin
memberships of the two members become uncorrelated. Its size is of order the number of bins over the
number of galaxies, negligible compared with $\xi$ but not compared with the leakage of Eq.~(\ref{eq:dxi}), as shown in
Sect.~\ref{sec:valid}. For subsets normalised separately,
\citet[their Eqs.~3.6 and 3.7]{demattia19} include the corresponding
shot-noise contribution, which for a binned map reduces to
Eq.~(\ref{eq:dxi_poisson_binned}). Equation~(\ref{eq:dxi_poisson}) gives
it for any parametrisation. Pair-count estimators remove the average of this term over all
pairs, $-\sum_kf_k^2/\E[N_k]$ with $f_k$ the volume fraction of bin
$k$, through their normalisation by the number of distinct pairs
\citep{peebles80,ls93}. The part that survives in the measured
correlation function is
\begin{equation}
   \langle\Delta\xi_{\rm P}\rangle_r^{\rm norm} =
   -\sum_k\frac{f_{kk}(r)-f_k^2}{\E[N_k]}\,,
   \label{eq:dxi_poisson_norm}
\end{equation}
which is negative at small separations, where pairs share a bin more
often than the average pair does, and tends to zero where
$f_{kk}(r)\to f_k^2$. At intermediate separations it can change sign,
depending on the geometry of the regions.

\subsection{Computing the correction in a survey analysis}
\label{ssec:recipe}

In a survey, the correlation function, $\hat\xi$, is measured with the estimated
selection function $\Shat$. By Eq.~(\ref{eq:dxi}), what is measured is not the true $\xi$ but $\xi+\Delta\xi[\xi]$, where we write
$\Delta\xi[\xi]$ for the leakage computed from a given correlation
function. The leakage in the data is $\Delta\xi[\xi]$, which
requires the true $\xi$, the quantity one wants to measure.
Subtracting $\Delta\xi[\hat\xi]$ instead is possible only as an
approximation, which Sect.~\ref{sec:valid} tests.

As for the global integral constraint, the solution is to correct
the model rather than the data. When fitting, each trial correlation
function $\xi_{\rm model}$ is turned into what the survey would measure
if the truth were $\xi_{\rm model}$, namely
$\xi_{\rm model}+\Delta\xi[\xi_{\rm model}]$, which, projected onto
the measured statistic in the same way as the data, is compared with
$\hat\xi$. Since $\Delta\xi$ is linear in
$\xi$, the correction is a fixed linear operator acting on the model,
which can be computed once and then applied to every trial model
\citep[as in the integral-constraint formalism of][]{demattia19}.
In practice, with $\xi_{\rm model}$ tabulated in fine bins $s$
of separation, the correction in measurement bin $r$ is
\begin{equation}
   \langle\Delta\xi\rangle_r \simeq \sum_s \mM_{rs}\,\xi_{\rm model}(s)\,,
   \label{eq:Mij}
\end{equation}
where the response matrix $\mM$ depends only on the footprint, the
fitted selection function and the pair weighting, and is computed
once. The input bins $s$ extend to the largest separation in the
survey, well beyond the measured bins $r$. Each likelihood evaluation
then applies $\mM$ to the trial model.

For binned completeness, the model tested in
Sect.~\ref{sec:valid}, the computation takes four steps. The leakage at
each pair is Eq.~(\ref{eq:dxi_binned}), which depends only on the
regions $A_k$ and on $\xi$, through the region averages $\bar\xi_k$
and $\bar\xi_{kl}$, and not on the completeness $p_k$ of the bins,
which cancel between $\bg$ and $\mH^{-1}$. First, each master
random is assigned to its bin of the systematics, which defines the
regions $A_k$ as sets of randoms. Second, the region averages are
computed from the trial correlation function. They are volume averages, so they
are computed with the master randoms, unweighted, whatever the pair
weighting of the estimator. With $N_{{\rm rand},k}$ the number of
master randoms in $A_k$,
\begin{align}
   \bar\xi_k(\by) &\simeq \frac{1}{N_{{\rm rand},k}}
      \sum_{a\in A_k}\xi_{\rm model}(|\bx_a-\by|)\,,
   \notag\\
   \bar\xi_{kl} &\simeq \frac{\sum_s RR_{kl}(s)\,\xi_{\rm model}(s)}
                              {\sum_s RR_{kl}(s)}\,,
   \label{eq:recipe_binned}
\end{align}
where $RR_{kl}(s)$ are the unweighted random pair counts between $A_k$ and $A_l$
in fine bins of separation $s$, extending to the largest separation in
the survey. The first is the mean correlation between a point $\by$
and region $A_k$, the second the mean correlation between two
regions. Third, Eq.~(\ref{eq:dxi_binned}), with these averages, gives the
leakage for every pair of master randoms. Fourth, the leakage is
averaged over the pairs in each separation bin with the weights of
the estimator,
\begin{equation}
   \langle\Delta\xi\rangle_r \simeq
   \frac{\sum_{(a,b)\in r}w_{ab}\,\Delta\xi(\bx_a,\bx_b)}
        {\sum_{(a,b)\in r}w_{ab}}\,,
   \label{eq:recipe_pairavg}
\end{equation}
where the sums run over the pairs of master randoms with separation
in bin $r$, and $w_{ab}$ is the weight with which a pair at those
positions enters the estimator: $w_{ab}=1$ if
the galaxies are weighted by $1/\Shat$ and the randoms are uniform,
and $w_{ab}=\Shat(\bx_a)\Shat(\bx_b)$ if the galaxies are unweighted
and the randoms are weighted or downsampled by $\Shat$. The
completeness values enter only here, and not at all when $w=1$. The
shot-noise term of Eq.~(\ref{eq:dxi_poisson_norm}) is added, with
$f_{kk}(r)$ computed from the same pairs.

For a general parametrisation, the leakage of Eq.~(\ref{eq:dxi})
is built from $\bg$, $\mH$, $\ba$, and $\mK$. The first is known
pointwise once $S$ is fitted. The other three are integrals over the
footprint, which we evaluate as sums over the master random catalogue
$\mathcal R$ with Eq.~(\ref{eq:rand_identity}):
\begin{align}
   \mH &\simeq \frac{\nb}{\nrand}\sum_{a\in\mathcal R}
          S(\bx_a)\,\bg(\bx_a)\,\bg(\bx_a)^T\,,
   \notag\\
   \ba(\by) &\simeq \frac{\nb}{\nrand}\sum_{a\in\mathcal R}
          \nabla_\theta S(\bx_a)\,\xi_{\rm model}(|\bx_a-\by|)\,,
   \label{eq:recipe_general}\\
   \mK &\simeq \Bigl(\frac{\nb}{\nrand}\Bigr)^2\sum_{a\neq b}
          \nabla_\theta S(\bx_a)\,\nabla_\theta^TS(\bx_b)\,
          \xi_{\rm model}(|\bx_a-\bx_b|)\,.
   \notag
\end{align}
The factors of $\nb$ cancel in $\mH^{-1}\ba$ and $\mH^{-1}\mK\mH^{-1}$,
so the result does not depend on the convention for $\nb$. Inserting
these in Eq.~(\ref{eq:dxi}) gives $\Delta\xi(\bx_a,\bx_b)$ for every
pair of randoms, and Eq.~(\ref{eq:recipe_pairavg}) averages it
over the pairs in each bin as before. In both cases the response
matrix $\mM$ of Eq.~(\ref{eq:Mij}) follows by applying these steps to
each input bin $s$ of $\xi_{\rm model}$ in turn.

Two practical points follow. First, $\xi_{\rm model}$ is needed at all
the separations spanned by the regions $A_k$, which for scattered
regions can be most of the survey, and not only in the range where
$\hat\xi$ is measured. Section~\ref{ssec:valid_results} shows that
truncating it biases the result. Second, the same integrals give
the sample-variance part of the error on the completeness estimate,
$\mK$ of Eq.~(\ref{eq:recipe_general}). For the binned model the
region averages $\bar\xi_{kl}$ of Eq.~(\ref{eq:recipe_binned}) give
Eq.~(\ref{eq:pk_var}) directly. Computing the correction also provides realistic errors on the completeness map, which Poisson
statistics underestimate.

\section{Numerical validation}
\label{sec:valid}

We now test the predictions of Sects.~\ref{sec:variance} and
\ref{sec:leakage} in controlled conditions. We impose a known
selection function on simulated halo catalogues, estimate it from the
selected catalogues as a survey would, and compare what the estimate
does with what the theory predicts. We use the binned model, for
which the predictions are explicit and the true selection function
is exactly representable, so that any discrepancy would point to the
theory rather than to a mismatch between model and truth. Four
predictions are tested:
\begin{enumerate}
   \item in each catalogue, the fitted completeness of each bin is
         shifted by the mean overdensity of its region
         (Eq.~\ref{eq:pk_sv});
   \item over catalogues, its variance is the Poisson term plus the
         variance of the mean overdensity of the region
         (Eq.~\ref{eq:pk_var});
   \item the correlation function of the corrected catalogue is
         offset by the leakage of Eq.~(\ref{eq:dxi_binned});
   \item without clustering, only the shot-noise term of
         Eq.~(\ref{eq:dxi_poisson_norm}) remains.
\end{enumerate}
The first three use the halo catalogues of FastPM simulations
\citep{feng16}. The fourth uses unclustered catalogues of the same
density, in which there is no overdensity to absorb. In all of
them the catalogue before selection is known, so the true mean
overdensity of each region is available for comparison.

\begin{figure}
\centering
\includegraphics[width=\columnwidth]{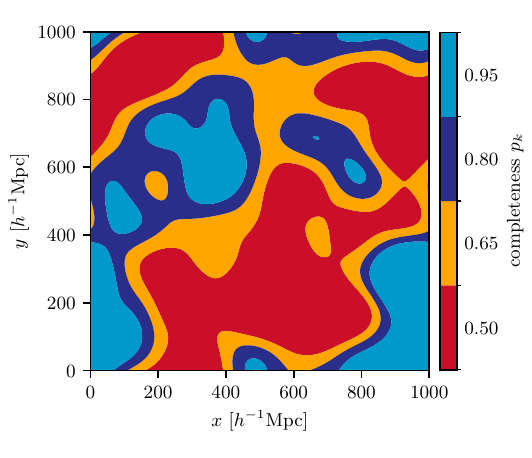}
\caption{Slice through the middle of the box showing the four
completeness regions $A_k$ used in the tests, the quartiles of the
systematic map, with the completeness $p_k$ of each. The map is the
same for all catalogues.}
\label{fig:regions}
\end{figure}

\begin{figure*}
\centering
\includegraphics[width=\textwidth]{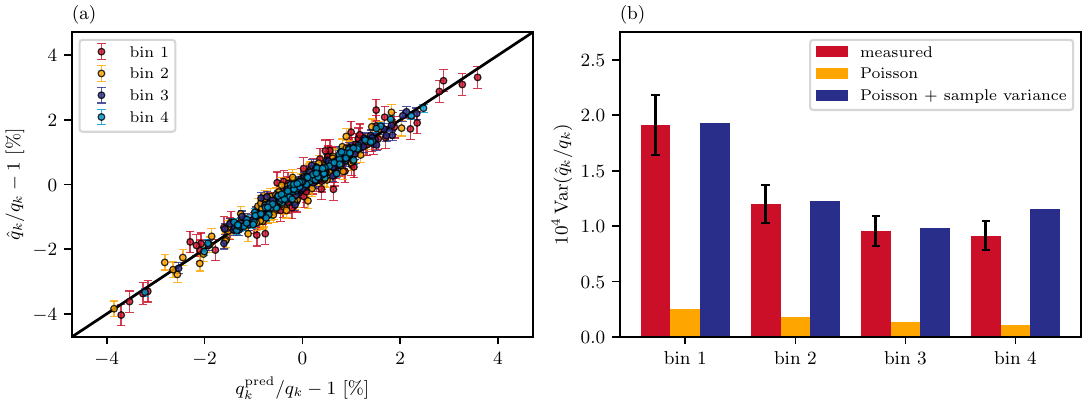}
\caption{Completeness estimates on the 100 halo catalogues. (a) Shift of the fitted completeness of each bin,
$\hat q_k/q_k-1$, against the shift predicted from the mean
overdensity of the region the bin occupies, $q^{\rm pred}_k/q_k-1$; each point is one bin of one catalogue, and the
error bars are the binomial scatter due to the random selection alone.
(b) Variance of $\hat q_k/q_k$ over the catalogues
(red; error bars: $\sqrt{2/99}$ of the measured variance, the
sampling error of a variance estimated from 100 catalogues), compared with the Poisson term (orange) and with
the prediction of Eq.~(\ref{eq:pk_var}) (blue), which adds to the
Poisson term the variance $\bar\xi_{kk}$ of the mean overdensity of
the region; both are normalised like $\hat q_k$.}
\label{fig:completeness}
\end{figure*}

\begin{figure*}
\centering
\includegraphics[width=\textwidth]{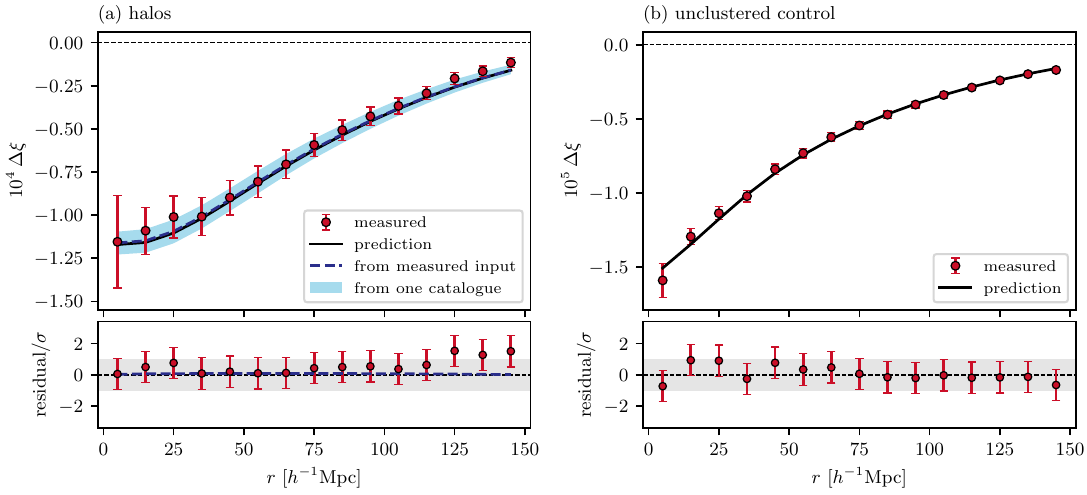}
\caption{Leakage into the correlation function, measured as
$\hat\xi(\Shat)-\hat\xi(S)$ averaged over the catalogues.
(a) Halos: measurement (points) and the prediction of
Sect.~\ref{ssec:recipe} computed with the true $\xi$ (solid line) and with the
mean over the catalogues of $\hat\xi$ measured with $\Shat$ (dashed
line). The cyan band is the 68\% range, over the catalogues, of
the prediction computed with the $\hat\xi$ of each catalogue in turn,
as a single survey would. All inputs are given in the same bins of
$2.5\,h^{-1}$Mpc and pass through the same linear operator.
(b) Unclustered control: measurement
and the shot-noise term of Eq.~(\ref{eq:dxi_poisson_norm}). The
lower panels show the difference between the measurement and the
solid-line prediction in units of the error bar of the measurement
(grey band: $\pm1\sigma$); in (a) the dashed line is the same for
the prediction from the measured input.}
\label{fig:leakage}
\end{figure*}

\subsection{Catalogues and measurements}
\label{ssec:valid_setup}

\begin{table}
\caption{Simulated catalogues.}
\label{tab:setup}
\centering
\small
\begin{tabular}{@{}ll@{}}
\hline\hline
Simulations & 100 FastPM boxes \\
Cosmology & Planck 2018 \\
Box side & $1000\,h^{-1}$Mpc \\
Particles & $672^3$ \\
Time steps & 10, equally spaced in $a$ \\
Redshift range & 9 to 1 \\
Halo finder & FoF, linking length 0.2 \\
Minimum halo size & 20 particles \\
Halo density & $2.65\times10^{-4}\,h^3{\rm Mpc}^{-3}$ \\
\hline
\end{tabular}
\end{table}

We use 100 FastPM simulations \citep{feng16}, a particle-mesh scheme
that reproduces the large-scale results of N-body codes with few time
steps, in the Planck 2018 cosmology \citep{planck18}, with the
settings of Table~\ref{tab:setup}. Halos are identified
at $z=1$ with a friends-of-friends algorithm \citep{davis85}, and we take all
halos with at least 20 particles. As a control we generate 1000 catalogues of
uniformly distributed points with the same mean density, which have
no clustering.

On these catalogues we impose a selection function driven by a
single systematic map $s(\bx)$, playing the role of an observing
condition. The map is a Gaussian random field, independent of the
halos, with power spectrum $P_s(k)\propto\exp(-k^2R^2/2)$ and
$R=100\,h^{-1}$Mpc, so that it varies on scales of about $R$. We
rank-transform it so that its values are uniformly distributed
between 0 and 1, which leaves its spatial pattern unchanged and makes
each quarter of the range of $s$ cover a quarter of the volume. The
selection function is then a step function of $s$. We divide $s$ into
its four quartiles and assign them the completeness
$p=(0.50,0.65,0.80,0.95)$, from the worst to the best observing
conditions. Each object is kept with the completeness of the bin in
which it falls, and discarded otherwise.
The binned model of Sect.~\ref{ssec:stepfn} is the correct model, and the tests isolate the effect of sample variance from model
misspecification. Figure~\ref{fig:regions} shows the resulting regions
$A_k$ on a slice of the box. Each covers a quarter of the volume in
many patches, a few hundred $h^{-1}$Mpc across. The same map, and so the same regions, is used to modulate the visibility of all
catalogues, together with a single master random catalogue with roughly 20
times as many points as a catalogue before selection.

On each contaminated catalogue we fit the completeness with
Eq.~(\ref{eq:pk_nobs}). The fit fixes it only up to an overall
constant, which Eq.~(\ref{eq:pk_nobs})
chooses so that the fitted completeness averages to one over the
volume. When comparing true and fitted values we rescale the true completeness in the same way, $q_k\equiv p_k/\sum_lf_lp_l$
with $f_l$ the volume fraction of bin $l$, and write $\hat q_k$ for
the fitted values. For our four bins of equal volume, the true values
above become $q_k=0.69$, 0.90, 1.10, and 1.31.

We then measure
$\hat\xi(r)$ with the
\textsc{MeasCorr} pair-count code \citep{farina26}, weighting the halos once by $1/\Shat$ and once by the true
$1/S$. The pair counts use
the analytic random pairs of the periodic box, which corresponds to the
$w=1$ pair weighting of Eq.~(\ref{eq:dxi_r}). The measured leakage is
the difference $\hat\xi(\Shat)-\hat\xi(S)$ in 15 bins of width
$10\,h^{-1}$Mpc, which we compare with the predictions computed for
the same bins, that is, averaged over the pairs whose separation falls
in each bin, Eq.~(\ref{eq:dxi_r}).

The four tests listed at the start of this section differ in the
information, the fitted completeness, its scatter over
catalogues, or the correlation function, their predictions require. The first, the shift of the
fitted completeness of each bin, concerns a single catalogue and
needs the actual overdensity of each region in it,
which a survey cannot know but a simulation can:
$\bar\delta_k=N_k^{\rm true}/(\nb V_k)-1$, with $N_k^{\rm true}$ the
number of objects in the region before the selection.
Eq.~(\ref{eq:pk_sv}) then gives the expected fitted completeness
$q^{\rm pred}_k$, normalised like $\hat q_k$. The other three, the variance of the
completeness (Eq.~\ref{eq:pk_var}), the leakage
(Eq.~\ref{eq:dxi_binned}, computed as in Sect.~\ref{ssec:recipe}), and
the shot-noise term (Eq.~\ref{eq:dxi_poisson_norm}), are
statistical predictions for averages and variances over catalogues.
They need only the systematic map and a correlation function, which in
an analysis of real data would be the model being fitted. Unless stated otherwise, we use
instead the true $\xi$ of the halos, so that the test probes the
correction rather than the modelling of $\xi$: the average of $\xi$
measured with \textsc{MeasCorr} on the first 50 of the same
simulations, before the selection is applied, the same
for all catalogues.
The prediction integrates the input over pairs of points anywhere in
two regions (Sect.~\ref{ssec:recipe}), so it needs $\xi$ at all the
separations the regions span, not only where $\hat\xi$ is measured.
We tabulate it in bins of $2.5\,h^{-1}$Mpc, interpolate linearly
between the bin centres, and continue it beyond $150\,h^{-1}$Mpc, up to the largest separation in
the periodic box (about $870\,h^{-1}$Mpc), by $b^2\xi_{\rm lin}$, with $b^2$ fitted between 60 and
$140\,h^{-1}$Mpc and $\xi_{\rm lin}$ the linear correlation function
at $z=1$ computed with CLASS \citep{lesgourgues11,blas11} for the
cosmology of the simulations. The resulting prediction is then
averaged over the $10\,h^{-1}$Mpc bins of the measurement (Eq.~\ref{eq:dxi_r}), as for any model compared with binned data.
In Sect.~\ref{ssec:valid_results} we also compute the leakage
prediction with $\hat\xi$ in place of the true $\xi$: the correlation
function measured with \textsc{MeasCorr} on the contaminated
catalogues, weighting the halos by $1/\Shat$, in the same
$2.5\,h^{-1}$Mpc bins, interpolated and continued beyond
$150\,h^{-1}$Mpc in the same way as the model.

\subsection{Results}
\label{ssec:valid_results}

The first prediction concerns the completeness estimate itself: in
each catalogue, the fitted completeness of each bin absorbs the mean
overdensity $\bar\delta_k$ of its region (Eq.~\ref{eq:pk_sv}).

Figure~\ref{fig:completeness}a shows this for each bin of each
catalogue: the shift found by the fit, $\hat q_k/q_k-1$, against the
shift expected from the true mean overdensity of the region,
$q^{\rm pred}_k/q_k-1$ (Sect.~\ref{ssec:valid_setup}). The error bars
are the binomial scatter due to the random selection alone. The points follow the diagonal closely. In every catalogue the fit absorbs the mean overdensity of each
region, as predicted.

In a survey the overdensity of a region is not known, so the same
shift is an error of the completeness estimate. It differs from
catalogue to catalogue, and its variance adds to the Poisson noise.
The second prediction (Eq.~\ref{eq:pk_var}) is that the variance of $\hat q_k$ over catalogues is the Poisson term
$1/\E[N_k]$ plus the variance $\bar\xi_{kk}$ of the mean overdensity
of the region, a quantity that depends only on $\xi$ and on the shape
of the region. Since $\hat q_k$ is normalised to unit mean, we
normalise the prediction in the same way, which removes from both
terms the part common to all bins. Figure~\ref{fig:completeness}b
compares both with the variance measured over the 100 catalogues,
bin by bin. The measured variance is six to nine times the Poisson
value, and Eq.~(\ref{eq:pk_var}) predicts it
within its statistical uncertainty, about 14\% for a
variance estimated from 100 catalogues, in all bins but
the fourth, which shows a fluctuation of about $1.5\sigma$.

The third prediction is the central one. Weighting each halo by the
inverse of the fitted completeness removes, together with the
selection, the mean overdensity of each region, so that $\hat\xi$ is
lower than with the true completeness by the leakage of
Eq.~(\ref{eq:dxi_binned}). Figure~\ref{fig:leakage}a compares it with the
measurement.
The points are the measured leakage of Sect.~\ref{ssec:valid_setup},
averaged over the catalogues. In each catalogue we measure $\hat\xi$
twice from the same halos, weighting them once with the fitted and
once with the true completeness, so that the difference
$\hat\xi(\Shat)-\hat\xi(S)$ is due to the fit alone. The error
bars are the uncertainty of the mean over the 100 catalogues. The solid line is the
theoretical prediction, the leakage of Eq.~(\ref{eq:dxi_binned})
computed as in Sect.~\ref{ssec:recipe} with the true $\xi$. It has no
free parameter, since its inputs, the systematic map and the true $\xi$, are fixed independently of the leakage measurement. The measured difference is negative at all separations and varies
slowly, from about $10^{-4}$ at $r<40\,h^{-1}$Mpc to $10^{-5}$ at
$145\,h^{-1}$Mpc, while $\xi$ falls by three orders of magnitude. As Eq.~(\ref{eq:dxi_binned}) anticipates, the leakage is set by averages
of $\xi$ over the regions, not by $\xi$ at the separation $r$. The
prediction agrees with the measurement at all separations, as the residuals in the lower
panel show. The
neglected second-order terms, of order $\xi(r)\,\bar\xi$, could become
visible where $\xi(r)$ is not small, and are here below the
statistical errors.

Two points matter when the prediction is applied to a survey. First, as anticipated in Sect.~\ref{ssec:valid_setup}, an input correlation function is needed out to the size of the regions, far beyond the
separations at which $\hat\xi$ is usually measured. Setting the halo $\xi$ to zero beyond
$150\,h^{-1}$Mpc, which removes its negative tail, raises
$\bar\xi_{kk}$ by 50 to $70\%$. Most of this change is common to all
regions and cancels in the leakage, which becomes more negative by
only $1\%$ at small separations, but by $25\%$ at
$145\,h^{-1}$Mpc, more than the statistical error there.
Second, in a real survey analysis the true $\xi$ is not known, and the prediction is computed either from the model $\xi$ of the fit, as in
Sect.~\ref{ssec:recipe}, or from the measured $\hat\xi$, which
already contains the leakage. The dashed line shows the prediction
computed with the mean over the catalogues of $\hat\xi$ measured with
$\Shat$, so that the two predictions differ only by their input.
A survey has a single $\hat\xi$, so we also compute the
prediction with the $\hat\xi$ of each catalogue separately. The cyan
band shows the $68\%$ range of these predictions over the
catalogues.
Since the prediction is linear in its input, computing it with $\hat\xi=\xi+\Delta\xi$ instead of $\xi$ changes it by the leakage
of the leakage, $\Delta\xi[\Delta\xi]$, which is smaller than
$\Delta\xi[\xi]$ by the ratio $\Delta\xi/\xi$ on the separations that
dominate $\bar\xi_{kl}$, about $1\%$ here. Accordingly, the dashed
line is $1\%$ less negative than the solid one. The difference would grow where the leakage is a larger
fraction of $\xi$: for weakly clustered tracers, or where
$\bar\xi_{kk}$ is larger, i.e. with finer bins of the systematics,
whose regions $A_k$ are smaller, and is absent when the model $\xi$
is used. The half-width of the cyan band
is $6\%$ of the prediction at $5\,h^{-1}$Mpc and $17\%$ at
$145\,h^{-1}$Mpc. In our set-up, a survey that computes the
correction from its own $\hat\xi$ therefore recovers it to within
this fraction.

The fourth, and final, prediction that we test is the shot-noise term
of Sect.~\ref{ssec:shotnoise}. Even without clustering, the number of
objects found in each region fluctuates around its expectation, and
the fitted completeness of the bin absorbs this Poisson fluctuation
as it absorbs the overdensity of the region. Correcting the catalogue
with it propagates the shot noise of the counts into $\hat\xi$, and,
since the same objects then form the pairs, it biases $\hat\xi$ instead of only adding noise. For the halos this term is about an
order of magnitude smaller than the leakage and below the statistical
errors of the 100 simulations. We therefore test it on the 1000
unclustered control catalogues, where there is no overdensity to
absorb and $\hat\xi(\Shat)-\hat\xi(S)$ contains only this term. Figure~\ref{fig:leakage}b shows this test. The points are
$\hat\xi(\Shat)-\hat\xi(S)$ averaged over the control catalogues,
with error bars again the uncertainty of the mean, and
the solid line is the theoretical prediction
(Eq.~\ref{eq:dxi_poisson_norm}).
The measured term is largest at small separations and decreases with
$r$ as fewer pairs fall within a single region. It is clearly
different from zero, and the prediction follows it at all
separations (lower panel).

\section{An application to real data: BOSS DR12 CMASS}
\label{sec:boss}

As a demonstrative example of the size of the effect in a real
analysis, we compute the leakage for the CMASS galaxies of BOSS DR12
\citep{dawson13,alam15}. The example is
meant only to illustrate the method on a survey whose selection
function was fitted on its own galaxies. We use the public large-scale-structure and random catalogues of \citet{reid16}, with the
selection of
\citet{cuesta16}: $0.43<z<0.7$, about $7.8\times10^5$ galaxies in
the northern and southern Galactic caps (NGC and SGC), and the
weights of the official catalogues, including FKP weights
\citep{fkp94} with $P_0=2\times10^4\,h^{-3}{\rm Mpc}^3$.

We use the selection function of the official analysis, which
corrects for the density of foreground stars and for the seeing
\citep{ross11,ross12,reid16}: Eqs.~(32) and (33) of \citet{ross17}. This raises two issues that we address in Appendix~\ref{app:ee} and ~\ref{app:marks}. First, their selection function minimises a
$\chi^2$ of binned counts rather than the Poisson likelihood of Eq.~(\ref{eq:composite}). Appendix~\ref{app:ee} shows that the
leakage of such a fit follows from the same derivation, and that on
the BOSS counts the $\chi^2$ and the likelihood fit give leakages
within $0.5\%$ of each other (Fig.~\ref{fig:appC}). Second, the
correction for the stellar density is fitted in bins of the fibre
magnitude of the galaxies, their magnitude within the fibre aperture,
a property that correlates with the clustering. Brighter galaxies are
more strongly biased, so that each bin absorbs the overdensity of a
subsample with its own bias.
Appendix~\ref{app:marks} extends the derivation to this case and
gives the details of the computation.

\begin{figure*}
\centering
\includegraphics[width=\textwidth]{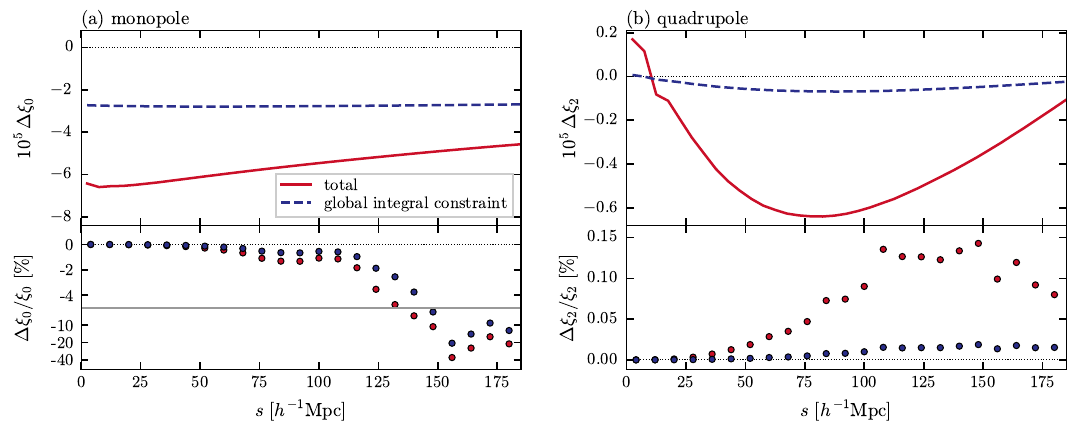}
\caption{Difference between the correlation function measured on
BOSS DR12 CMASS galaxies ($0.43<z<0.7$, NGC and SGC combined) and the
true one, for the monopole (a) and the quadrupole (b): the total
(red), sum of the global integral constraint
of the measurement and of the leakage of the fitted selection
function, and the global integral constraint alone (blue).
Top: the difference. Bottom: its ratio to the multipoles
measured by \citet{cuesta16}. For the monopole the vertical scale
is linear up to $5\%$ (grey line) and logarithmic beyond.}
\label{fig:boss}
\end{figure*}

The measurement also carries the global integral constraint (Eq.~\ref{eq:ic_single}) from the normalisation of the random
catalogue of each cap to its galaxies, which is present whether $S$ is
fitted or not. Figure~\ref{fig:boss} shows both terms, computed with
Eq.~(\ref{eq:recipe_pairavg}) and the FKP weights of the measurement
and combined over the two caps with their pair counts, as the
measurement does.

The top row of Fig.~\ref{fig:boss} shows the
difference between the measured and the true multipoles. The total
(red), sum of the global integral constraint and of the leakage of
the fit, and the constraint alone (blue). In the monopole, the global integral constraint is nearly constant,
$-2.8\times10^{-5}$, and fitting the selection function roughly
doubles it. The total is $-6.5\times10^{-5}$ at $20\,h^{-1}$Mpc and
$-4.6\times10^{-5}$ at $180\,h^{-1}$Mpc, of which the constraint
accounts for 42 to $58\%$. As in Sect.~\ref{sec:valid}, the
leakage decreases slowly with separation, being set by averages of
$\xi$ over the regions the fit acts on and not by $\xi$ at the separation $s$. In the quadrupole the constraint is small and the
leakage dominates, nearly ten times larger, because the patterns
that the fit absorbs, on the sky and in redshift, are not isotropic.
The total reaches $-0.64\times10^{-5}$ at $80\,h^{-1}$Mpc and
decreases towards zero at the largest separations.

The bottom row compares the difference, with the same colours,
with the multipoles measured by \citet{cuesta16}. Since the monopole decreases with $s$
while the difference is nearly constant, their ratio grows with
separation, from $0.3\%$ at $50\,h^{-1}$Mpc to $1\%$ at
$100\,h^{-1}$Mpc and about $40\%$ just beyond $150\,h^{-1}$Mpc,
where the measured $\xi_0$ approaches zero. In the quadrupole the ratio stays at about
$0.1\%$.

\section{Summary and discussion}
\label{sec:conc}

We have treated the estimation of the survey selection function
as the maximisation of the Poisson likelihood of unclustered
galaxies, applied to a clustered catalogue, and asked how the
estimate changes the correlation function measured after the
correction. A broad class
of estimators, from counts in bins of a systematic to templates and
neural networks, maximise this
likelihood (Eq.~\ref{eq:composite}),  and
differ only in how $S$ is parametrised. . Because the catalogue is clustered, the fit absorbs part of
the large-scale structure. The estimate is noisy not only through
shot noise but also through the structure it absorbs, and correcting the
catalogue removes that structure from the measured correlation
function. In general, the correction imposes a generalised integral
constraint, one per parameter, weighted by the sensitivity of $S$ to
that parameter.
For a binned completeness map the weights are the regions occupied by
the bins, and each bin absorbs the mean overdensity of its region.

The additional noise of the estimate has a counterpart in
spatial statistics. For clustered point processes, the covariance of
intensity estimates is the sandwich formula of
\citet{waagepetersen07,waagepetersen09}, and Eq.~(\ref{eq:sandwich})
is the same result with the clustering set by the correlation
function of the tracers. Survey analyses usually treat the errors on
completeness maps as Poisson and so miss this term, which in our
tests on simulated catalogues makes the variance six to nine times
the Poisson value. The
effect of the fit on the clustering itself (Eq.~\ref{eq:dxi}) has
to our knowledge no counterpart in that literature, where the
intensity is the quantity of interest and the clustering a
nuisance.

In cosmology, \citet{demattia19} showed that estimating the
radial selection and the overall density from the data, or
normalising the density in sky pixels, imposes integral
constraints, which they model as a linear operator acting on
the model correlation function. For imaging systematics, the eBOSS and DESI analyses
estimate the corresponding effect numerically, by applying the same weighting procedure to mocks and measuring its effect on the mock
correlation function \citep{demattia21,chaussidon25,chiarenza26}.
Equation~(\ref{eq:dxi}) extends the formalism of \citet{demattia19} to any selection
function estimated from the data, whatever its parametrisation, and
gives in analytic form the effect that those analyses estimate
from mocks. The global integral constraint is its simplest case, a
selection function constant over the survey (a single bin).

The template literature contains the same effect in another
form. \citet{elsner16,elsner17} derived the clustering bias of
template subtraction and mode projection, and its analytic
correction, for additive contamination, and \citet{weaverdyck21}
extended them to multiplicative contamination at next order. For
log-linear templates, $\ln S=\theta^T\bt$, Eq.~(\ref{eq:dxi}) with
$\bg=\bt$ gives the corresponding leakage, the subtraction from
$\delta$ of its $S$-weighted projection onto the templates. The same
equation covers the other methods compared by \citet{weaverdyck21},
ordinary and penalised regression, the neural networks of
\citet{rezaie21}, and, through Appendix~\ref{app:ee}, the
mean-squared-error fits of \citet{rezaie20} and the $\chi^2$ fits of
binned galaxy-to-random ratios. All estimate $S$ from the data and differ only
in $\bg$, $\mH$ and, for penalised fits, the penalty.

The Poisson noise of the estimate adds the shot-noise term of
Eq.~(\ref{eq:dxi_poisson}), of order the number of bins over the
number of galaxies for a binned map, small for coarse bins in a
large survey, but measurable in our control catalogues. For flexible
models with many parameters this term and the leakage both grow.
Together they are the mechanism usually described as overfitting
\citep{rezaie20,weaverdyck21}, whose bias \citet{rezaie21} found to be
almost linear in the true power, as Eq.~(\ref{eq:dxi}) implies.

On FastPM halo catalogues with a known selection function, and on
unclustered control catalogues, the binned predictions are reproduced
with no free parameter. In each catalogue the completeness estimate
absorbs the mean overdensity of its region, its variance exceeds the
Poisson value by the variance of that mean overdensity, and the
leakage into the correlation function and the shot-noise term match
the predictions. The uncertainty of the completeness
and the correction of the clustering must therefore account for the
fluctuations absorbed by the fit, even when the selection model
represents the imposed selection exactly.

We recommend including the leakage in the model rather than
correcting the data. Since the leakage is linear in the true
correlation function, it can be computed once with the random
catalogue as a linear operator (Eq.~\ref{eq:Mij}). In the fit, this
operator is applied to each trial model, and the result is added to
the model before comparison with the data. The operator requires the
correlation function over the separations spanned by the completeness
regions, well beyond the fitted range. In our set-up, truncating this
input changes the correction appreciably at large separations.
Computing it from the measured $\hat\xi$ instead of the model adds
the leakage of the leakage, about $1\%$ in our tests. This fraction
is set by the ratio of the leakage to $\xi$, and would be larger for
weakly clustered tracers or finer bins of the systematics.

To illustrate the size of the effect on a real catalogue, we
computed the leakage of the official stellar-density and seeing
weights of BOSS DR12 CMASS into the measured correlation function.
Applying the method to these weights required two extensions. First, the BOSS weights
are a $\chi^2$ fit of binned galaxy-to-random ratios, not a Poisson
maximum-likelihood estimate. Appendix~\ref{app:ee} shows that the
derivation of the leakage only uses the fact that the fitted
parameters make a weighted count of the galaxies equal to the count
expected from the selection function, with weights that depend on the estimator. The $\chi^2$ fit, like any estimator of this form,
leaks in the same way as the maximum-likelihood fit, with its own
weights in place of the sensitivity of $S$. Second, the
correction for the stellar density is fitted in bins of the fibre
magnitude of the galaxies, a property that correlates with their bias. Appendix~\ref{app:marks} treats the catalogue as a marked point
process, in which each magnitude bin absorbs the overdensity of its
own galaxies and the randoms carry marks drawn from the distribution
of the galaxies. In this example, the leakage of the BOSS weights is about
$0.5\%$ of the monopole at the scale of the baryon acoustic peak,
$1\%$ together with the global integral constraint of the
measurement, which is small for a BAO analysis. It grows to over
$40\%$ at larger separations, where the monopole approaches zero,
and is therefore more important for measurements dominated by the
largest scales, such as constraints on primordial non-Gaussianity.\\
The treatment assumes that detection depends on known observing
conditions, and possibly on properties of the galaxy
(Appendix~\ref{app:marks}), and is independent between galaxies at
fixed selection probability. Density-dependent detection and unmodelled selection
variables are excluded. The results are first order in $\delta$ and
$\Delta\theta$, except where noted exact for the binned model. The
second-order terms, of order $\xi\bar\xi$, are below the statistical
errors in our tests and grow with finer bins of the systematics.
Finally, the
leakage is the price of estimating the selection function from the
data being corrected, and only that. If $S$ is calibrated externally,
from source injection \citep{suchyta16,everett22,kong20} or from a
reference sample whose clustering does not enter, then $\bb=0$, and
there is neither a sample-variance term nor leakage.\\
The natural next steps are to apply the correction within a
full survey analysis, from the fit of the selection function to the
cosmological constraints, and to propagate the fit of the selection function into the
covariance of the clustering measurement. These applications will show when the analytic
correction can replace or complement the one estimated from mocks,
and how flexible a selection model can be before it removes
cosmological information.

$$$$
\begin{acknowledgements}Simulations
and computations in this work have been run at the computing facilities of
INFN, Sezione di Genova: the authors wish to thank the INFN IT personnel
in Genova for their precious and constant support.
AF and AV acknowledge support from the INAF "Astrofisica di Frontiera 2025"
Curiosity Science grant SFIDA (Statistical Framework for Informed
Data-driven Analysis of galaxy clustering; CUP C73C26000490005).
AF and EB acknowledge support from the MIUR PRIN 2022 grant 2022NY2ZRS001 ``Optimizing the extraction of cosmological information from Large Scale Structure analysis in view of the next large spectroscopic surveys''and from the ASI/INAF agreement n.~2024-10-HH.0
\end{acknowledgements}

\bibliographystyle{aa}

\bibliography{refs}

@ARTICLE{kids1000,
  author = {Asgari, M. and Lin, C.-A. and Joachimi, B. and others},
  title = {{KiDS-1000 cosmology: Cosmic shear constraints and comparison between two point statistics}},
  journal = {A\&A}, year = 2021, volume = 645, pages = {A104}
}

@ARTICLE{bianchi17,
  author = {Bianchi, D. and Percival, W. J.},
  title = {{Unbiased clustering estimation in the presence of missing observations}},
  journal = {MNRAS}, year = 2017, volume = 472, pages = {1106}
}

@ARTICLE{campbell09,
  author = {Campbell, N. R.},
  title = {{The study of discontinuous phenomena}},
  journal = {Proc. Cambridge Philos. Soc.}, year = 1909, volume = 15, pages = {117--136}
}

@ARTICLE{chaussidon22,
  author = {Chaussidon, E. and Y{\`e}che, C. and Palanque-Delabrouille, N. and others},
  title = {{Angular clustering properties of the DESI QSO target selection using DR9 Legacy Imaging Surveys}},
  journal = {MNRAS}, year = 2022, volume = 509, pages = {3904}
}

@BOOK{daley03,
  author = {Daley, D. J. and Vere-Jones, D.},
  title = {{An Introduction to the Theory of Point Processes. Vol. I: Elementary Theory and Methods}}, edition = {2nd},
  publisher = {Springer}, address = {New York}, year = 2003
}

@ARTICLE{davis82,
  author = {Davis, M. and Huchra, J.},
  title = {{A survey of galaxy redshifts. III. The density field and the induced gravity field}},
  journal = {ApJ}, year = 1982, volume = 254, pages = {437}
}

@ARTICLE{dawson13,
  author = {Dawson, K. S. and Schlegel, D. J. and Ahn, C. P. and others},
  title = {{The Baryon Oscillation Spectroscopic Survey of SDSS-III}},
  journal = {AJ}, year = 2013, volume = 145, pages = {10}
}

@ARTICLE{dawson16,
  author = {Dawson, K. S. and Kneib, J.-P. and Percival, W. J. and others},
  title = {{The SDSS-IV extended Baryon Oscillation Spectroscopic Survey: Overview and early data}},
  journal = {AJ}, year = 2016, volume = 151, pages = {44}
}

@ARTICLE{demattia19,
  author = {de Mattia, A. and Ruhlmann-Kleider, V.},
  title = {{Integral constraints in spectroscopic surveys}},
  journal = {JCAP}, year = 2019, volume = 2019, number = {08}, pages = {036}
}

@ARTICLE{demattia21,
  author = {de Mattia, A. and Ruhlmann-Kleider, V. and Raichoor, A. and others},
  title = {{The completed SDSS-IV extended Baryon Oscillation Spectroscopic Survey: measurement of the BAO and growth rate of structure of the emission line galaxy sample from the anisotropic power spectrum between redshift 0.6 and 1.1}},
  journal = {MNRAS}, year = 2021, volume = 501, pages = {5616}
}

@ARTICLE{chaussidon25,
  author = {Chaussidon, E. and Y{\`e}che, C. and de Mattia, A. and others},
  title = {{Constraining primordial non-Gaussianity with DESI 2024 LRG and QSO samples}},
  journal = {JCAP}, year = 2025, volume = 2025, number = {06}, pages = {029}
}

@ARTICLE{chiarenza26,
  author = {Chiarenza, S. and Krolewski, A. and Bonici, M. and others},
  title = {{Constraining primordial non-Gaussianity from DESI DR1 quasars and Planck PR4 CMB lensing}},
  journal = {JCAP}, year = 2026, volume = 2026, number = {05}, pages = {098}
}

@ARTICLE{spergel15,
  author = {Spergel, D. and Gehrels, N. and Baltay, C. and others},
  title = {{Wide-Field InfrarRed Survey Telescope-Astrophysics Focused Telescope Assets WFIRST-AFTA 2015 Report}},
  journal = {arXiv e-prints}, year = 2015, pages = {arXiv:1503.03757}
}

@ARTICLE{euclid25,
  author = {{Euclid Collaboration: Mellier}, Y. and Abdurro'uf and others},
  title = {{Euclid. I. Overview of the Euclid mission}},
  journal = {A\&A}, year = 2025, volume = 697, pages = {A1}
}

@ARTICLE{des3yr,
  author = {{DES Collaboration} and Abbott, T. M. C. and Aguena, M. and others},
  title = {{Dark Energy Survey Year 3 results: Cosmological constraints from galaxy clustering and weak lensing}},
  journal = {Phys. Rev. D}, year = 2022, volume = 105, pages = {023520}
}

@ARTICLE{desi16,
  author = {{DESI Collaboration} and Aghamousa, A. and Aguilar, J. and others},
  title = {{The DESI Experiment Part I: Science, targeting, and survey design}},
  journal = {arXiv e-prints}, year = 2016, pages = {arXiv:1611.00036}
}

@ARTICLE{elsner16,
  author = {Elsner, F. and Leistedt, B. and Peiris, H. V.},
  title = {{Unbiased methods for removing systematics from galaxy clustering measurements}},
  journal = {MNRAS}, year = 2016, volume = 456, pages = {2095}
}

@ARTICLE{elsner17,
  author = {Elsner, F. and Leistedt, B. and Peiris, H. V.},
  title = {{Unbiased pseudo-$C_\ell$ power spectrum estimation with mode projection}},
  journal = {MNRAS}, year = 2017, volume = 465, pages = {1847}
}

@ARTICLE{farina26,
  author = {Farina, A. and Veropalumbo, A. and Branchini, E. and Guidi, M.},
  title = {{Modeling and measuring the anisotropic halo 3-point correlation function: a coordinated study}},
  journal = {JCAP}, year = 2026, volume = 2026, number = {02}, pages = {028}
}

@ARTICLE{feng16,
  author = {Feng, Y. and Chu, M.-Y. and Seljak, U. and McDonald, P.},
  title = {{FastPM: a new scheme for fast simulations of dark matter and haloes}},
  journal = {MNRAS}, year = 2016, volume = 463, pages = {2273}
}

@ARTICLE{hikage19,
  author = {Hikage, C. and Oguri, M. and Hamana, T. and others},
  title = {{Cosmology from cosmic shear power spectra with Subaru Hyper Suprime-Cam first-year data}},
  journal = {PASJ}, year = 2019, volume = 71, pages = {43}
}

@ARTICLE{kong20,
  author = {Kong, H. and Burleigh, K. J. and Ross, A. and others},
  title = {{Removing imaging systematics from galaxy clustering measurements with Obiwan: application to the SDSS-IV extended Baryon Oscillation Spectroscopic Survey emission-line galaxy sample}},
  journal = {MNRAS}, year = 2020, volume = 499, pages = {3943}
}

@ARTICLE{layzer56,
  author = {Layzer, D.},
  title = {{A new model for the distribution of galaxies in space}},
  journal = {AJ}, year = 1956, volume = 61, pages = {383}
}

@ARTICLE{ls93,
  author = {Landy, S. D. and Szalay, A. S.},
  title = {{Bias and variance of angular correlation functions}},
  journal = {ApJ}, year = 1993, volume = 412, pages = {64}
}

@BOOK{peebles80,
  author = {Peebles, P. J. E.},
  title = {{The Large-Scale Structure of the Universe}},
  publisher = {Princeton University Press}, address = {Princeton}, year = 1980
}

@ARTICLE{pinol17,
  author = {Pi{\~n}ol, N. and Cahn, R. N. and Hand, N. and Seljak, U. and White, M.},
  title = {{Imprint of DESI fiber assignment on the anisotropic power spectrum of emission line galaxies}},
  journal = {JCAP}, year = 2017, volume = 2017, number = {04}, pages = {008}
}

@ARTICLE{rezaie20,
  author = {Rezaie, M. and Seo, H.-J. and Ross, A. J. and Bunescu, R. C.},
  title = {{Improving galaxy clustering measurements with deep learning: analysis of the DECaLS DR7 data}},
  journal = {MNRAS}, year = 2020, volume = 495, pages = {1613}
}

@ARTICLE{rezaie21,
  author = {Rezaie, M. and Ross, A. J. and Seo, H.-J. and others},
  title = {{Primordial non-Gaussianity from the completed SDSS-IV extended Baryon Oscillation Spectroscopic Survey I: catalogue preparation and systematic mitigation}},
  journal = {MNRAS}, year = 2021, volume = 506, pages = {3439}
}

@ARTICLE{rosadomarin25,
  author = {Rosado-Mar{\'i}n, A. J. and Ross, A. J. and Seo, H.-J. and others},
  title = {{Mitigating imaging systematics for DESI 2024 emission line galaxies and beyond}},
  journal = {JCAP}, year = 2025, volume = 2025, number = {07}, pages = {018}
}

@ARTICLE{ross11,
  author = {Ross, A. J. and Ho, S. and Cuesta, A. J. and others},
  title = {{Ameliorating systematic uncertainties in the angular clustering of galaxies: a study using the SDSS-III}},
  journal = {MNRAS}, year = 2011, volume = 417, pages = {1350}
}

@ARTICLE{waagepetersen07,
  author = {Waagepetersen, R.},
  title = {{An estimating function approach to inference for inhomogeneous Neyman-Scott processes}},
  journal = {Biometrics}, year = 2007, volume = 63, pages = {252--258}
}

@ARTICLE{waagepetersen09,
  author = {Waagepetersen, R. and Guan, Y.},
  title = {{Two-step estimation for inhomogeneous spatial point processes}},
  journal = {J. R. Stat. Soc. B}, year = 2009, volume = 71, pages = {685--702}
}

@ARTICLE{weaverdyck21,
  author = {Weaverdyck, N. and Huterer, D.},
  title = {{Mitigating contamination in LSS surveys: a comparison of methods}},
  journal = {MNRAS}, year = 2021, volume = 503, pages = {5061}
}

@ARTICLE{ross12,
  author = {Ross, A. J. and Percival, W. J. and S{\'a}nchez, A. G. and others},
  title = {{The clustering of galaxies in the SDSS-III Baryon Oscillation Spectroscopic Survey: analysis of potential systematics}},
  journal = {MNRAS}, year = 2012, volume = 424, pages = {564}
}

@ARTICLE{groth77,
  author = {Groth, E. J. and Peebles, P. J. E.},
  title = {{Statistical analysis of catalogs of extragalactic objects. VII. Two- and three-point correlation functions for the high-resolution Shane-Wirtanen catalog of galaxies}},
  journal = {ApJ}, year = 1977, volume = 217, pages = {385}
}

@BOOK{moller04,
  author = {M{\o}ller, J. and Waagepetersen, R. P.},
  title = {{Statistical Inference and Simulation for Spatial Point Processes}},
  publisher = {Chapman \& Hall/CRC}, address = {Boca Raton}, year = 2004
}

@ARTICLE{schoenberg05,
  author = {Schoenberg, F. P.},
  title = {{Consistent parametric estimation of the intensity of a spatial-temporal point process}},
  journal = {J. Stat. Plan. Inference}, year = 2005, volume = 128, pages = {79--93}
}

@ARTICLE{johnston21,
  author = {Johnston, H. and Wright, A. H. and Joachimi, B. and others},
  title = {{Organised randoms: Learning and correcting for systematic galaxy clustering patterns in KiDS using self-organising maps}},
  journal = {A\&A}, year = 2021, volume = 648, pages = {A98}
}

@ARTICLE{yan25,
  author = {Yan, Z. and Wright, A. H. and Chisari, N. E. and others},
  title = {{KiDS-Legacy: Angular galaxy clustering from deep surveys with complex selection effects}},
  journal = {A\&A}, year = 2025, volume = 694, pages = {A259}
}

@ARTICLE{cox55,
  author = {Cox, D. R.},
  title = {{Some statistical methods connected with series of events}},
  journal = {J. R. Stat. Soc. B}, year = 1955, volume = 17, pages = {129--157}
}

@ARTICLE{melchior21,
  author = {Melchior, P. and Joseph, R. and Sanchez, J. and MacCrann, N. and Gruen, D.},
  title = {{The challenge of blending in large sky surveys}},
  journal = {Nat. Rev. Phys.}, year = 2021, volume = 3, pages = {712--718}
}

@ARTICLE{euclid26xcii,
  author = {{Euclid Collaboration: Monaco}, P. and Elkhashab, M. and others},
  title = {{Euclid preparation. XCII. Controlling angular systematics in the Euclid spectroscopic galaxy sample}},
  journal = {A\&A}, year = 2026, volume = 712, pages = {A165}
}

@ARTICLE{suchyta16,
  author = {Suchyta, E. and Huff, E. M. and Aleksi{\'c}, J. and others},
  title = {{No galaxy left behind: accurate measurements with the faintest objects in the Dark Energy Survey}},
  journal = {MNRAS}, year = 2016, volume = 457, pages = {786}
}

@ARTICLE{everett22,
  author = {Everett, S. and Yanny, B. and Kuropatkin, N. and others},
  title = {{Dark Energy Survey Year 3 Results: Measuring the Survey Transfer Function with Balrog}},
  journal = {ApJS}, year = 2022, volume = 258, pages = {15}
}

@ARTICLE{planck18,
  author = {{Planck Collaboration} and Aghanim, N. and Akrami, Y. and others},
  title = {{Planck 2018 results. VI. Cosmological parameters}},
  journal = {A\&A}, year = 2020, volume = 641, pages = {A6}
}

@ARTICLE{davis85,
  author = {Davis, M. and Efstathiou, G. and Frenk, C. S. and White, S. D. M.},
  title = {{The evolution of large-scale structure in a universe dominated by cold dark matter}},
  journal = {ApJ}, year = 1985, volume = 292, pages = {371}
}

@ARTICLE{blas11,
  author = {Blas, D. and Lesgourgues, J. and Tram, T.},
  title = {{The Cosmic Linear Anisotropy Solving System (CLASS). Part II: Approximation schemes}},
  journal = {JCAP}, year = 2011, volume = 2011, number = {07}, pages = {034}
}

@ARTICLE{lesgourgues11,
  author = {Lesgourgues, J.},
  title = {{The Cosmic Linear Anisotropy Solving System (CLASS) I: Overview}},
  journal = {arXiv e-prints}, year = 2011, pages = {arXiv:1104.2932}
}

@INPROCEEDINGS{belghazi18,
  author = {Belghazi, M. I. and Baratin, A. and Rajeshwar, S. and others},
  title = {{Mutual information neural estimation}},
  booktitle = {Proceedings of the 35th International Conference on Machine Learning},
  series = {PMLR}, volume = 80, year = 2018, pages = {531--540}
}

@ARTICLE{neyman58,
  author = {Neyman, J. and Scott, E. L.},
  title = {{Statistical approach to problems of cosmology}},
  journal = {J. R. Stat. Soc. B}, year = 1958, volume = 20, pages = {1--29}
}

@ARTICLE{alam15,
  author = {Alam, S. and Albareti, F. D. and Allende Prieto, C. and others},
  title = {{The Eleventh and Twelfth Data Releases of the Sloan Digital Sky Survey: Final Data from SDSS-III}},
  journal = {ApJS}, year = 2015, volume = 219, pages = {12}
}

@ARTICLE{reid16,
  author = {Reid, B. and Ho, S. and Padmanabhan, N. and others},
  title = {{SDSS-III Baryon Oscillation Spectroscopic Survey Data Release 12: galaxy target selection and large-scale structure catalogues}},
  journal = {MNRAS}, year = 2016, volume = 455, pages = {1553}
}

@ARTICLE{ross17,
  author = {Ross, A. J. and Beutler, F. and Chuang, C.-H. and others},
  title = {{The clustering of galaxies in the completed SDSS-III Baryon Oscillation Spectroscopic Survey: observational systematics and baryon acoustic oscillations in the correlation function}},
  journal = {MNRAS}, year = 2017, volume = 464, pages = {1168}
}

@ARTICLE{cuesta16,
  author = {Cuesta, A. J. and Vargas-Maga{\~n}a, M. and Beutler, F. and others},
  title = {{The clustering of galaxies in the SDSS-III Baryon Oscillation Spectroscopic Survey: baryon acoustic oscillations in the correlation function of LOWZ and CMASS galaxies in Data Release 12}},
  journal = {MNRAS}, year = 2016, volume = 457, pages = {1770}
}

@ARTICLE{alam17,
  author = {Alam, S. and Ata, M. and Bailey, S. and others},
  title = {{The clustering of galaxies in the completed SDSS-III Baryon Oscillation Spectroscopic Survey: cosmological analysis of the DR12 galaxy sample}},
  journal = {MNRAS}, year = 2017, volume = 470, pages = {2617}
}

@ARTICLE{beutler17,
  author = {Beutler, F. and Seo, H.-J. and Saito, S. and others},
  title = {{The clustering of galaxies in the completed SDSS-III Baryon Oscillation Spectroscopic Survey: anisotropic galaxy clustering in Fourier space}},
  journal = {MNRAS}, year = 2017, volume = 466, pages = {2242}
}

@ARTICLE{fkp94,
  author = {Feldman, H. A. and Kaiser, N. and Peacock, J. A.},
  title = {{Power-spectrum analysis of three-dimensional redshift surveys}},
  journal = {ApJ}, year = 1994, volume = 426, pages = {23}
}

@ARTICLE{kaiser87,
  author = {Kaiser, N.},
  title = {{Clustering in real space and in redshift space}},
  journal = {MNRAS}, year = 1987, volume = 227, pages = {1}
}

@INPROCEEDINGS{neyman49,
  author = {Neyman, J.},
  title = {{Contribution to the theory of the $\\chi^2$ test}},
  booktitle = {Proceedings of the Berkeley Symposium on Mathematical Statistics and Probability},
  editor = {Neyman, J.},
  publisher = {University of California Press},
  address = {Berkeley},
  year = {1949},
  pages = {239--273}
}

@ARTICLE{baker84,
  author = {Baker, S. and Cousins, R. D.},
  title = {{Clarification of the use of chi-square and likelihood functions in fits to histograms}},
  journal = {Nuclear Instruments and Methods in Physics Research},
  volume = {221},
  pages = {437--442},
  year = {1984}
}

\begin{appendix}

\section{Derivation of the likelihood}
\label{app:lik}

This appendix derives Eq.~(\ref{eq:ppp_lik}) from the two
defining properties of a Poisson process stated in
Sect.~\ref{ssec:ppp_lik}.
Divide $\W$ into $M$ small cells $C_j$, of volume $\Delta V_j$ and
centred on $\bx_j$. We take the cells small enough that
$\lambda_{\rm obs}$ is nearly constant within each one, so that its
expected count is
\begin{equation}
   \mu_j=\int_{C_j}\lambda_{\rm obs}(\bx;\theta)\,\dd\bx
   \simeq\lambda_{\rm obs}(\bx_j;\theta)\,\Delta V_j\,,
   \label{eq:cell_mean}
\end{equation}
and small enough that $\mu_j\ll1$, so that the probability of finding
two or more galaxies in the same cell is negligible. By property (i),
the count $N_j$ in cell $j$ then takes only the values 0 and 1, with
probabilities
\begin{equation}
   P(N_j=0)=e^{-\mu_j}\,,\qquad P(N_j=1)=\mu_j\,e^{-\mu_j}\,.
   \label{eq:cell_prob}
\end{equation}
At this resolution a catalogue is fully specified by which cells are occupied. The $N$ cells containing the galaxies $\bx_1,\dots,\bx_N$
are occupied and the other $M-N$ are empty. By property (ii) the
probability of this configuration is the product of the cell
probabilities,
\begin{align}
   P(\{\bx_i\}\,|\,\theta,\nb,\delta)
   &=\prod_{i=1}^{N}\mu_i e^{-\mu_i}\prod_{j\ {\rm empty}}e^{-\mu_j}
   \notag\\
   &=\left(\prod_{i=1}^{N}\mu_i\right)\exp\!\left(-\sum_{j=1}^{M}\mu_j\right)\,,
   \label{eq:ppp_cells}
\end{align}
where $\mu_i$ denotes the mean of the cell containing galaxy $i$. In
the second line we used the fact that every cell, occupied or empty,
contributes one factor $e^{-\mu_j}$, so that all of them combine into
a single exponential of the sum over all $M$ cells, while only the
occupied cells contribute an extra factor $\mu_i$.

We now let the cells shrink. The sum in the exponent is the expected
number of galaxies in the whole footprint,
$\sum_{j=1}^M\mu_j=\int_\W\lambda_{\rm obs}\,\dd\bx$, which does not
depend on how finely the footprint is divided, since refining the cells
only splits the same total into more, smaller pieces, so $M$ drops
out. The other factor has one term per galaxy whatever $M$ is, and
each $\mu_i=\lambda_{\rm obs}(\bx_i;\theta)\,\Delta V_i$ by
Eq.~(\ref{eq:cell_mean}). The factors $\Delta V_i$ are the sizes of
the cells, which depend neither on $\theta$ nor on $\nb$. Dividing by
them turns the probability into a probability density in the galaxy
positions, which is the likelihood of Eq.~(\ref{eq:ppp_lik}).

\section{Regularised fits}
\label{app:reg}

Flexible models, such as the neural network of
Sect.~\ref{ssec:nn}, have many parameters that the data constrain
only in part, and are often regularised by a penalty on the
parameters. For a quadratic penalty of curvature $\mR$, centred on a
point $\theta_{\rm p}$, the fit no longer sets the score to zero but
to $\mR(\hat\theta-\theta_{\rm p})$, and the expansion of
Sect.~\ref{sec:variance} gives
\begin{equation}
   \Delta\theta \simeq \mA\,\bU(\theta_0) - \mA\,\mR\,(\theta_0-\theta_{\rm p})\,,
   \qquad \mA\equiv(\mH+\mR)^{-1}\,.
   \label{eq:dtheta_reg}
\end{equation}
The second term is a fixed shift of the estimate, independent of
the density field and of the shot noise. It drops out of every
covariance below, and acts on $\hat\xi$ as a misspecification of
$\Shat$ by a fixed fraction, not as leakage. The first term has the form of Eq.~(\ref{eq:dtheta})
with $\mH^{-1}$ replaced by $\mA$, while the score $\bU(\theta_0)$ is
the same as before, with covariance $\mH+\mK$. The results of
Sects.~\ref{sec:variance} and \ref{sec:leakage} carry over as
follows. The covariance of the estimate (Eq.~\ref{eq:sandwich})
becomes
\begin{equation}
   \Cov(\hat\theta)\simeq\mA\,(\mH+\mK)\,\mA\,.
   \label{eq:sandwich_reg}
\end{equation}
The leakage (Eq.~\ref{eq:dxi}) keeps its form with $\mH^{-1}$
replaced by $\mA$ in all three terms, since the clustering covariance
$\mK$ appears in it explicitly. The shot-noise term (Eq.~\ref{eq:dxi_poisson}) becomes
\begin{equation}
   \Delta\xi_{\rm P}(\bx,\by)=
   \bg(\bx)^T\bigl(-2\mA+\mA\,\mH\,\mA\bigr)\,\bg(\by)\,,
   \label{eq:dxi_poisson_reg}
\end{equation}
where the first part comes from the cross terms of
Eq.~(\ref{eq:delta_hat_P}) and the second from the variance of the
fitted selection function. All three results reduce to the
unregularised ones for $\mR=0$. Early stopping is not a fixed penalty,
and its response requires a treatment of its own.

\section{Fits that do not maximise the Poisson likelihood}
\label{app:ee}

Not every estimator of the selection function maximises
Eq.~(\ref{eq:composite}). The data-to-random ratio in bins of a
systematic is often fitted by $\chi^2$ \citep{ross12}, and the
networks of \citet{rezaie20} minimise a mean-squared error. This
appendix shows that the results of Sects.~\ref{sec:variance} and
\ref{sec:leakage} hold for these fits as well, with a change of
weights. We consider any fit that solves an estimating equation of
the form
\begin{equation}
   \bPsi(\theta)=\sum_i\bc(\bx_i;\theta)
   -\nb\int_\W\bc(\bz;\theta)\,S(\bz;\theta)\,\dd\bz=0\,,
   \label{eq:ee}
\end{equation}
where $\bc$ is a vector field with one component per parameter,
the weight with which a galaxy at $\bz$ enters the fit. By Campbell's
theorem the sum over galaxies has expectation
$\nb\int_\W(1+\delta)\,S\,\bc\,\dd\bz$, so $\bPsi$ vanishes on
average at $\theta_0$ for an unclustered catalogue, so the fit recovers
the true selection function in the absence of clustering. The score
of Eq.~(\ref{eq:score_def}) is the case $\bc=\bg$.

A $\chi^2$ fit has this form. Divide the footprint into regions
$B_j$, the bins of one or more systematics, with galaxy counts $N_j$
and expected counts $\mu_j(\theta)=\nb\int_{B_j}S\,\dd\bz$.
Minimising $\sum_j(N_j-\mu_j)^2/\sigma_j^2$ with errors $\sigma_j$
that do not depend on $\theta$ sets
$\sum_j(N_j-\mu_j)\nabla_\theta\mu_j/\sigma_j^2$ to zero, which is
Eq.~(\ref{eq:ee}) with
\begin{equation}
   \bc(\bz)=\frac{\nabla_\theta\mu_j}{\sigma_j^2}\,,
   \qquad \bz\in B_j\,,
   \label{eq:c_chi2}
\end{equation}
since $N_j$ is a sum over the galaxies in $B_j$ and
$\sum_j\mu_j\nabla_\theta\mu_j/\sigma_j^2=\nb\int_\W\bc\,S\,\dd\bz$.
Fitting the galaxy-to-random ratio $N_j/(\alpha R_j)$ instead of the
count $N_j$ gives the same $\chi^2$, with $\sigma_j$ multiplied by
$\alpha R_j$, and a mean-squared error on pixel counts is the case of
equal $\sigma_j$.

We now repeat the expansion of Sect.~\ref{sec:variance}.
Differentiating Eq.~(\ref{eq:ee}) with respect to $\theta$ at
$\theta_0$ and averaging at $\delta=0$, the derivatives of $\bc$ cancel between the
two terms, as the second derivatives of $\ln S$ did in
Eq.~(\ref{eq:fisher}), and minus the slope of $\bPsi$ is
\begin{equation}
   \mJ=\nb\int_\W S(\bz;\theta_0)\,\bc(\bz;\theta_0)\,
   \bg^T(\bz;\theta_0)\,\dd\bz\,,
   \label{eq:J_ee}
\end{equation}
which reduces to $\mH$ for $\bc=\bg$. To first order, $\Delta\theta\simeq\mJ^{-1}\bPsi(\theta_0)$,
and as in Eq.~(\ref{eq:dtheta}) the estimating function at the truth
splits into its mean at fixed density field,
$\bb_c(\delta)=\nb\int_\W\delta\,S\,\bc\,\dd\bz$, and a Poisson
fluctuation, of covariance $\mH_c=\nb\int_\W S\,\bc\,\bc^T\dd\bz$.
The fit therefore absorbs the overdensity averaged with weights
$S\,\bc$ rather than $\nabla_\theta S$. The corrected field is still
$\hat\delta\simeq\delta-\bg^T\Delta\theta$, since the correction
applies $S(\hat\theta)$ whatever the fit. With
\begin{equation}
   \begin{aligned}
   \ba_c(\by)&=\nb\int_\W S(\bz)\,\bc(\bz)\,\xi(\bz-\by)\,\dd\bz
\,,\\
   \mK_c&=\nb^2\iint_\W S(\bx)\,\bc(\bx)\,S(\by)\,\bc^T(\by)\,
   \xi(\bx-\by)\,\dd\bx\,\dd\by\,,
   \end{aligned}
   \label{eq:aK_ee}
\end{equation}
the counterparts of Eqs.~(\ref{eq:a_def}) and (\ref{eq:K}),
the steps that led to Eqs.~(\ref{eq:sandwich}), (\ref{eq:dxi})
and (\ref{eq:dxi_poisson}) give
\begin{equation}
   \Cov(\hat\theta)\simeq\mJ^{-1}(\mH_c+\mK_c)\,\mJ^{-T}\,,
   \label{eq:sandwich_ee}
\end{equation}
\begin{equation}
   \begin{aligned}
   \Delta\xi(\bx,\by) ={}&
   -\,\bg(\bx)^{T}\mJ^{-1}\ba_c(\by)
   -\,\bg(\by)^{T}\mJ^{-1}\ba_c(\bx)\\
   &+\,\bg(\bx)^{T}\mJ^{-1}\mK_c\,\mJ^{-T}\bg(\by)\,,
   \end{aligned}
   \label{eq:dxi_ee}
\end{equation}
\begin{equation}
   \begin{aligned}
   \Delta\xi_{\rm P}(\bx,\by) ={}&
   -\,\bg(\bx)^{T}\mJ^{-1}\bc(\by)
   -\,\bg(\by)^{T}\mJ^{-1}\bc(\bx)\\
   &+\,\bg(\bx)^{T}\mJ^{-1}\mH_c\,\mJ^{-T}\bg(\by)\,,
   \qquad \bx\neq\by\,.
   \end{aligned}
   \label{eq:dxi_poisson_ee}
\end{equation}
For $\bc=\bg$ these reduce to Eqs.~(\ref{eq:sandwich}),
(\ref{eq:dxi}) and (\ref{eq:dxi_poisson}), and a quadratic penalty
adds $\mR$ to $\mJ$, as in Appendix~\ref{app:reg}. The leakage is
still linear in $\xi$ and is computed with the random catalogue as in
Sect.~\ref{ssec:recipe}, with $S\,\bc$ in place of $\nabla_\theta S$
in $\ba$ and $\mK$ and with $\mJ$ in place of $\mH$.

The weights of BOSS DR12 CMASS \citep{reid16,ross17}, the
example of Sect.~\ref{sec:boss}, are of this kind. Two observing
conditions modulate the density of the galaxies: the density of
foreground stars, $n_\star$, and the $i$-band seeing, $S_i$. The
selection function is the product
\begin{equation}
   S = \bigl[A(i_{\rm fib2})+B(i_{\rm fib2})\,n_\star\bigr]\;
       A_{\rm see}\Bigl[1-{\rm erf}\Bigl(\frac{S_i-B_{\rm see}}{\sigma_{\rm see}}\Bigr)\Bigr]\,,
   \label{eq:boss_S}
\end{equation}
where $A$ and $B$ are constant within each of five bins of the
fibre magnitude $i_{\rm fib2}$ in the fit and are interpolated between
the bin centres when the weights are assigned. The two factors are
fitted separately on the same galaxies, over $0.35<z<1.0$, and
neither fit corrects for the other. The stellar-density factor in
each $i_{\rm fib2}$ bin with the two Galactic caps together, the
seeing factor in each cap. In both fits the galaxies and the randoms
are counted in ten bins of $n_\star$ or of $S_i$, and the model,
evaluated at the bin centres, is fitted to the galaxy-to-random ratio
of the bins with a $\chi^2$ whose errors are the Poisson errors of
the counts.\footnote{The fitting code is
public: \url{https://dr12.sdss.org/sas/dr12/boss/lss/mksampleDR12/}
(\texttt{mkweightfiles.py}).} The weights applied to the catalogue
are the inverse of Eq.~(\ref{eq:boss_S}).

In BOSS the errors of the $\chi^2$ are Poisson errors computed
from the observed counts, $\sigma_j^2=N_j$, kept fixed during the fit:
Neyman's $\chi^2$ \citep{neyman49}, whose relation to the
Poisson likelihood is discussed by \citet{baker84}. Its estimating
equation is
$\sum_j(N_j-\mu_j)\nabla_\theta\mu_j/N_j=0$. Maximising the Poisson
likelihood of the counts gives instead
$\sum_j(N_j-\mu_j)\nabla_\theta\mu_j/\mu_j=0$, the same equation
with the model counts $\mu_j(\theta)$ in place of the observed
counts $N_j$ in the weights. The two weights differ by
the residual $N_j-\mu_j$ that they multiply, so the two equations
differ at second order in the residuals. To first order the BOSS fit
is therefore the likelihood fit, whose
weight in Eq.~(\ref{eq:c_chi2}) is $\bc=\nabla_\theta\ln\mu_j$, the
mean of $\bg$ over $B_j$ weighted by $S$. For the step-function model
of Sect.~\ref{ssec:stepfn}, where $S$ is constant within each region,
this is $\bg$ itself, and Eqs.~(\ref{eq:sandwich}), (\ref{eq:dxi})
and (\ref{eq:dxi_poisson}) apply unchanged.

The BOSS fit also counts the galaxies in bins of the systematic
and evaluates the model at the bin centres, so that all the galaxies
of a bin enter the estimating equation with the same weight, the
value of $\bg$ at the centre of the bin, whereas in an unbinned fit
each galaxy enters with $\bg$ at its own position. The fit therefore
absorbs the overdensity averaged over each bin, while the weights it
returns are applied to each galaxy at its own value of the
systematic. Binning, a choice of the analysis rather than a property
of the $\chi^2$, changes which density fluctuations the fit absorbs,
and hence the size of the leakage, but not the way it is computed.

The second-order terms do shift the estimate. Their mean does not
vanish, because the squared residuals have positive mean,
$\mu_j+\mu_j^2\bar\xi_{jj}$, so that on average the fit moves the
selection function of each region by a fraction of order
$1/\mu_j+\bar\xi_{jj}$. Being of second order, this shift does not
contribute to the leakage, which is of first order in $\delta$. Its
mean, the same in every realisation, acts as a fixed error of the
weights.

\begin{figure*}
\centering
\includegraphics[width=\textwidth]{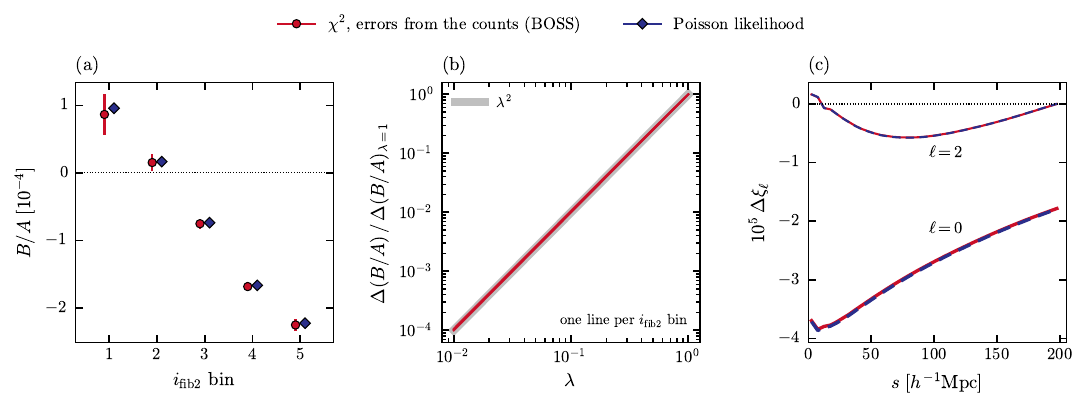}
\caption{The BOSS stellar-density fit of
Sect.~\ref{sec:boss} with two estimators of the same binned counts:
the $\chi^2$ with errors computed from the observed counts (the BOSS
analysis) and the Poisson likelihood. (a) Slope $B/A$ of the
stellar-density factor $A+B\,n_\star$ of Eq.~(\ref{eq:boss_S}), the
relative change of density per unit stellar density, in the five
$i_{\rm fib2}$ bins; the error bars are the
statistical errors of the $\chi^2$ fit. (b) Difference between the
$\chi^2$ and the likelihood estimates when the residuals of the
counts about the likelihood solution are scaled by $\lambda$,
normalised to its value at $\lambda=1$, for each of the five bins;
the grey line is $\lambda^2$. (c) Leakage of the two fits into the
measured monopole and quadrupole (shot-noise term included, global
integral constraint excluded), computed as in Sect.~\ref{sec:boss}.}
\label{fig:appC}
\end{figure*}

Figure~\ref{fig:appC} verifies these statements on the BOSS fit.
We repeat the stellar-density fit of each $i_{\rm fib2}$ bin on the
same binned galaxy and random counts by maximising their Poisson
likelihood, and compare the two through the ratio $B/A$, the relative change of the galaxy
density per unit stellar density, which is what the fit determines,
since $A$ only sets the normalisation of the bin. The slopes $B/A$ of
the $\chi^2$ differ from the likelihood estimates by $-0.10$ to
$-0.32$ times their statistical error (panel a). Panel b checks that this
difference is of second order in the residuals. We take the counts
expected by the likelihood fit, $\hat\mu_j$, and the residuals of the
data about them, $N_j-\hat\mu_j$, and we build counts with the same
residuals reduced by a factor $\lambda$,
$N_j(\lambda)=\hat\mu_j+\lambda(N_j-\hat\mu_j)$, which we fit with
both estimators. The likelihood fit returns the same parameters for
every $\lambda$, because its estimating equation is linear in the
counts. The difference between the two fits decreases as
$\lambda^2$ (panel b), as a quadratic term must. The same holds for the
parameters of the seeing factor, which differ between the two fits
by a few parts in $10^3$ and also scale as $\lambda^2$. The weights applied
to the galaxies, the inverse of Eq.~(\ref{eq:boss_S}) with the
parameters of each fit, differ between the two fits by $0.1\%$ for
a typical galaxy, and by less than $0.5\%$ for $99\%$ of the galaxies. The largest
differences, up to $3\%$, occur for the brightest galaxies at the
highest stellar densities. Panel c shows the leakage of the two
fits into the monopole and the quadrupole, computed as in
Sect.~\ref{sec:boss}. The two differ by less than $0.6\%$ of the
maximum of either.

\section{Selection functions that depend on galaxy properties}
\label{app:marks}

The selection function may also depend on a property of the
galaxy, a mark $m$ such as a magnitude or a colour. Then $S(\bx,m;\theta)$ is the detection probability of a galaxy with mark $m$ at $\bx$.
We assume that, at fixed mark, detection does not depend on the
density field, which is the independence assumption of
Sect.~\ref{ssec:thinning} applied to the galaxies of each mark. For
the sample as a whole, detection then does depend on the density,
because where the density is high the more strongly clustered marks
are over-represented. Since this dependence arises only through an
observed property of each galaxy, it is removed by following the
galaxies of each mark separately, with their own overdensity. Fibre
collisions, spectral overlaps and blending are different in kind. There the detection of a galaxy depends on the presence of other
galaxies nearby, that is, on the density itself, and they remain
excluded. The marked treatment is needed whenever the clustering
depends on the mark, and also whenever the fit has separate
parameters for galaxies of different marks, as the stellar-density
fit of BOSS in bins of fibre magnitude (Sect.~\ref{sec:boss}),
whether or not the clustering does. We now repeat the derivation of
Sects.~\ref{sec:lik}--\ref{sec:leakage} for this case.

The catalogue is a marked point process \citep{daley03}. Each galaxy has a position $\bx$ and a mark $m$. We write $\pi(m\,|\,\bx)$,
with $\int\pi(m\,|\,\bx)\,\dd m=1$, for the distribution of the marks
of the galaxies at $\bx$ before any selection, which in general depends
on position, for example through the redshift, and $\delta_m(\bx)$
for the overdensity of the galaxies with mark $m$. The intensity of
the true galaxy distribution on the space of positions and marks is
then $\nb\,\pi(m\,|\,\bx)\,[1+\delta_m(\bx)]$, and the overdensity of
all the galaxies is $\delta=\int\pi(m\,|\,\bx)\,\delta_m(\bx)\,\dd m$.
As in Sect.~\ref{ssec:pointproc}, given the fields $\delta_m$ the
galaxies are a Poisson process with this intensity on the space of
positions and marks. With the
independence assumption written as
$P(D=1\,|\,\bs,m,\{\delta_{m'}\})=P(D=1\,|\,\bs,m)$, thinning with the
probability $S(\bx,m;\theta)$ gives, in place of
Eq.~(\ref{eq:lambda_obs}),
\begin{equation}
   \lambda_{\rm obs}(\bx,m;\,\theta)=\nb\,\pi(m\,|\,\bx)\,[1+\delta_m(\bx)]\,S(\bx,m;\,\theta)\,.
   \label{eq:lambda_marks}
\end{equation}
The likelihood of Eq.~(\ref{eq:ppp_lik}) holds for the marked
process with $\lambda_{\rm obs}(\bx_i,m_i)$ at the galaxies and the
integral taken over positions and marks. Dropping the terms in
$\delta$ as in Eq.~(\ref{eq:composite}),
\begin{equation}
   \ln\mathcal{L}=N\ln\nb+\sum_i\ln S(\bx_i,m_i;\theta)
   +\sum_i\ln\pi(m_i\,|\,\bx_i)-\nb\,\Veff(\theta)\,,
   \label{eq:composite_marks}
\end{equation}
with $\Veff(\theta)=\int_\W\!\int S(\bx,m;\theta)\,\pi(m\,|\,\bx)\,\dd m\,\dd\bx$.
The term in $\pi$ does not depend on $\theta$, so the score is
Eq.~(\ref{eq:score_def}) with $\bg(\bx_i,m_i)$ at the galaxies and
$\nabla_\theta\Veff$ as the integral. Campbell's theorem (Eq.~\ref{eq:campbell}) holds on the space of positions and marks,
so by Eq.~(\ref{eq:lambda_marks}) the sum over galaxies has
expectation $\nb\int_\W\!\int(1+\delta_m)\,\nabla_\theta S\,\pi\,\dd m\,\dd\bx$,
and the bias of the score (Eq.~\ref{eq:score_bias}) becomes
\begin{equation}
   \bb(\delta)=\nb\int_\W\!\!\int\delta_m(\bx)\,\nabla_\theta S(\bx,m)\,\pi(m\,|\,\bx)\,\dd m\,\dd\bx\,.
   \label{eq:score_bias_marks}
\end{equation}
The fit absorbs the overdensity of the galaxies of each mark at each
position, with two weights: the sensitivity $\nabla_\theta S$ of the
selection function to the parameters at that mark, and the fraction
$\pi(m\,|\,\bx)$ of galaxies that carry it. The Fisher matrix (Eq.~\ref{eq:fisher}) contains no clustering and becomes
$\mH=\nb\int_\W\!\int S\,\bg\,\bg^T\,\pi\,\dd m\,\dd\bx$, and the
displacement of the estimate is Eq.~(\ref{eq:dtheta}) with these
$\bb$ and $\mH$. The variance of $\bb$ (Eq.~\ref{eq:K}) now involves
the correlation between the overdensities of two marks at two
points, $\langle\delta_m(\bx)\,\delta_{m'}(\by)\rangle=\xi_{mm'}(\bx-\by)$,
the cross-correlation function of the galaxies with marks $m$ and
$m'$,
\begin{equation}
   \begin{aligned}
   \mK&=\nb^2\iint_\W\!\!\iint\nabla_\theta S(\bx,m)\,\nabla_\theta^TS(\by,m')\,
   \pi(m\,|\,\bx)\,\pi(m'\,|\,\by)\\
   &\quad\times\xi_{mm'}(\bx-\by)\,\dd m\,\dd m'\,\dd\bx\,\dd\by\,,
   \end{aligned}
   \label{eq:K_marks}
\end{equation}
and Eq.~(\ref{eq:sandwich}) holds unchanged.

The clustering analysis weights each galaxy by
$1/\Shat(\bx_i,m_i)$. Dividing the observed density of galaxies
with mark $m$ by its expectation for an unclustered sample,
$\nb\,\pi(m\,|\,\bx)\,\Shat(\bx,m)$, as in Eq.~(\ref{eq:delta_hat_def}),
and expanding to first order as in Sect.~\ref{ssec:leak_general},
the corrected field of the galaxies with mark $m$ is
\begin{equation}
   \hat\delta(\bx,m)\simeq\delta_m(\bx)-\bg(\bx,m)^T\Delta\theta\,.
   \label{eq:delta_hat_marks}
\end{equation}
Its two-point function requires the cross-covariance between the
score bias and the overdensity of mark $m'$ at a second point, the
counterpart of Eq.~(\ref{eq:a_def}),
\begin{equation}
   \begin{aligned}
   \ba_{m'}(\by)&\equiv\langle\bb(\delta)\,\delta_{m'}(\by)\rangle\\
   &=\nb\int_\W\!\!\int\nabla_\theta S(\bz,m)\,\pi(m\,|\,\bz)\,\xi_{mm'}(\bz-\by)\,\dd m\,\dd\bz\,.
   \end{aligned}
   \label{eq:a_marks}
\end{equation}
Correlating Eq.~(\ref{eq:delta_hat_marks}) at two points, with
$\Delta\theta=\mH^{-1}\bb$ and $\langle\bb\,\bb^T\rangle=\mK$, gives
$\langle\hat\delta(\bx,m)\,\hat\delta(\by,m')\rangle=\xi_{mm'}(\bx-\by)+\Delta\xi(\bx,m;\by,m')$
with
\begin{equation}
   \begin{aligned}
   \Delta\xi(\bx,m;\by,m')={}&
   -\,\bg(\bx,m)^T\mH^{-1}\ba_{m'}(\by)
   -\,\bg(\by,m')^T\mH^{-1}\ba_{m}(\bx)\\
   &+\,\bg(\bx,m)^T\mH^{-1}\mK\,\mH^{-1}\bg(\by,m')\,,
   \end{aligned}
   \label{eq:dxi_marks}
\end{equation}
which is Eq.~(\ref{eq:dxi}) with the sensitivities evaluated at
the marks of the two galaxies and $\ba_{m'}$ in place of $\ba$. If the
weights applied to the galaxies are not the fitted model, as when the
parameters are interpolated in the mark when the weights are
assigned, $\bg$ in Eq.~(\ref{eq:delta_hat_marks}) is the gradient of
the applied weights, while $\mH$ follows the fitted model.

Equation~(\ref{eq:dxi_marks}) is the shift of the
cross-correlation of the galaxies with marks $m$ and $m'$. The
corrected field of the whole sample is the average of the
mark-resolved ones over the marks present at each position,
$\hat\delta(\bx)=\int\pi(m\,|\,\bx)\,\hat\delta(\bx,m)\,\dd m$, since
the density weighted by $1/\Shat$ is $\nb\,[1+\hat\delta(\bx)]$. Its
two-point function is therefore the average of
Eq.~(\ref{eq:dxi_marks}) with weights $\pi(m\,|\,\bx)\,\pi(m'\,|\,\by)$,
$\langle\hat\delta(\bx)\,\hat\delta(\by)\rangle=\xi(\bx-\by)+\Delta\xi(\bx,\by)$
with
\begin{equation}
   \begin{aligned}
   \Delta\xi(\bx,\by)={}&
   -\,\bar\bg(\bx)^T\mH^{-1}\bar\ba(\by)
   -\,\bar\bg(\by)^T\mH^{-1}\bar\ba(\bx)\\
   &+\,\bar\bg(\bx)^T\mH^{-1}\mK\,\mH^{-1}\bar\bg(\by)\,,\\
   \bar\bg(\bx)={}&\int\bg(\bx,m)\,\pi(m\,|\,\bx)\,\dd m\,,\\
   \bar\ba(\by)={}&\int\ba_{m'}(\by)\,\pi(m'\,|\,\by)\,\dd m'\,,
   \end{aligned}
   \label{eq:dxi_marks_avg}
\end{equation}
where the bar denotes here an average over the marks. By
Eq.~(\ref{eq:a_marks}), $\bar\ba$ is $\ba_{m'}$ with $\xi_{mm'}$ replaced by
$\int\xi_{mm'}\,\pi(m'\,|\,\by)\,\dd m'=\langle\delta_m(\bz)\,\delta(\by)\rangle$,
the cross-correlation of the galaxies of mark $m$ with the whole
sample. Pair-count estimators average Eq.~(\ref{eq:dxi_marks_avg})
over the pairs as in Eq.~(\ref{eq:dxi_r}).

The shot-noise term follows as in Sect.~\ref{ssec:shotnoise}. The
shot noise of the observed density is a white noise $\epsilon(\bx,m)$
on the space of positions and marks, with variance $\nb\,\pi S$,
since the observed galaxies are a Poisson process on that space. The
Poisson part of the corrected field is
\begin{equation}
   \hat\delta_{\rm P}(\bx,m)=\frac{\epsilon(\bx,m)}{\nb\,\pi(m\,|\,\bx)\,S(\bx,m)}
   -\bg(\bx,m)^T\mH^{-1}\!\int_\W\!\!\int\bg\,\epsilon\,\dd m'\,\dd\bz\,,
   \label{eq:delta_hat_P_marks}
\end{equation}
and the correlation of its two pieces at distinct points gives
\begin{equation}
   \Delta\xi_{\rm P}(\bx,m;\by,m')=-\,\bg(\bx,m)^T\mH^{-1}\bg(\by,m')\,,
   \label{eq:dxi_poisson_marks}
\end{equation}
which is Eq.~(\ref{eq:dxi_poisson}) at the marks of the two galaxies.
For the whole sample it is averaged over the marks as in
Eq.~(\ref{eq:dxi_marks_avg}), giving
$-\bar\bg(\bx)^T\mH^{-1}\bar\bg(\by)$, and the pair-count
normalisation removes its average over all pairs as in
Eq.~(\ref{eq:dxi_poisson_norm}).

Fits that do not maximise the likelihood are treated as in
Appendix~\ref{app:ee}. The weight field becomes $\bc(\bx,m)$, the
integrals are taken over positions and marks with the fraction $\pi$
as weight,
\begin{equation}
   \mJ=\nb\int_\W\!\!\int S\,\bc\,\bg^T\,\pi\,\dd m\,\dd\bx\,,
   \qquad
   \mH_c=\nb\int_\W\!\!\int S\,\bc\,\bc^T\,\pi\,\dd m\,\dd\bx\,,
   \label{eq:JH_marks}
\end{equation}
and $S\,\bc\,\pi$ replaces $\nabla_\theta S\,\pi$ in $\bb$, $\ba_m$
and $\mK$.

These expressions simplify if the mark changes only the amplitude
of the clustering, through a linear bias: $\xi_{mm'}=r_mr_{m'}\,\xi$,
where $r_m=b_m/\bar b$ is the bias of mark $m$ relative to a
reference bias $\bar b$ and $\xi$ is the correlation function at that
bias. Each point in Eqs.~(\ref{eq:K_marks}) and (\ref{eq:a_marks})
then carries the relative bias of its mark.
In the computation, the integrals over positions and marks are sums
over the random catalogue, as in Sect.~\ref{ssec:recipe}. For the
sum to sample the marks as well as the positions, each random is
given a mark $m$ drawn from $\pi(m\,|\,\bx)$, at which $S$, $\bg$ and
$\nabla_\theta S$ are evaluated. Since $\xi_{mm'}=r_mr_{m'}\,\xi$, in
$\ba_m$ and $\mK$ each random also carries its factor $r_m$, so that
a single $\xi$ serves for all pairs. In
Sect.~\ref{sec:boss} we take $\pi$ as the magnitude distribution of
the galaxies in bins of redshift, and one relative bias per magnitude
bin, neglecting the variation of the bias within it.

The relative biases $r_m$ of the magnitude bins are measured from
the monopole of the cross-correlation of each bin with the whole
sample between 20 and $80\,h^{-1}$Mpc, modelled with linear bias and
the Kaiser factor: $r_m=1.21$, 1.06, 1.05, 0.98 and 0.93 from the
brightest to the faintest bin, with mean one over the galaxies, as
the cross-correlations with the whole sample require. The
rescaling is applied to the whole redshift-space $\xi$, which also
rescales the terms due to peculiar velocities, although these do not
depend on the bias. The error of this approximation is a fraction of
the total effect of the relative biases. Repeating the computation
with the same bias for all the bins changes the leakage by about
$4\%$, so we expect it to be unimportant for the demonstrative example
of Sect.~\ref{sec:boss}.

The leakage of Sect.~\ref{sec:boss} is then computed as in
Sect.~\ref{ssec:recipe}, with the randoms of the fit domain,
$0.35<z<1.0$, on which the weights were fitted, for the integrals
over the footprint, and those of the measured sample, $0.43<z<0.7$,
for the pair averages, weighted with the FKP weights as in the
measurement. For the correlation function we use the linear Kaiser
model \citep{kaiser87} in the BOSS fiducial cosmology \citep{alam17}
at the effective redshift $z=0.6$, with the bias measured by
\citet{beutler17}. Being analytic, the model is evaluated directly at
every separation in the fit domain, and no extrapolation of the kind
used in Sect.~\ref{sec:valid} is needed.

\end{appendix}

\end{document}